\documentclass[article,twocolumn,nofootinbib,prd]{revtex4}
\usepackage[letterpaper,left=.75in,right=.75in,top=.75in,bottom=1.in]{geometry}
\usepackage{graphicx,amsfonts,color,comment,amsmath,hyperref,float}
\usepackage{amssymb}	
\usepackage{scrextend}
\usepackage{mathrsfs,amssymb}  
\usepackage{cancel}
\usepackage[normalem]{ulem}

\newcommand{\be}{\begin{equation}}
\newcommand{\ee}{\end{equation}}
\newcommand{\bea}{\begin{eqnarray}}
\newcommand{\eea}{\end{eqnarray}}

\begin{document}
%%%%%%%%%%%%%%%%%%%%%%%%%%%%%%%%%%%%%%%%%%%%%%%%%%%%%%%%%%%  FRONT PAGE
\title{Probing BSM Neutrino Physics with Flavor and Spectral Distortions:\\ Prospects for Future High-Energy Neutrino Telescopes}
%\title{Demystifying the Flavor of Cosmic Neutrinos}

\author{Ian M. Shoemaker}
\author{Kohta Murase}
\affiliation{Department of Physics; Department of Astronomy \& Astrophysics; Center for Particle and Gravitational Astrophysics, The Pennsylvania State University, University Park, PA 16802, USA}

\date{\today}
\begin{abstract}
{The flavor of cosmic neutrinos may help unveil their sources and could reveal the presence of new physics in the neutrino sector.  We consider impacts of next-generation neutrino detectors, including the planned upgrade to neutrino detector--{\it IceCube-Gen2 }, which is well-positioned to make dramatic improvements in both flavor and spectral measurements.  
We show that various models in neutrino physics beyond the Standard Model, such as neutrino decay, pseudo-Dirac states, and neutrino self-scattering, may be found or strongly constrained at {\it IceCube-Gen2 } and KM3NeT.  We find that the additional flavor discriminants given by Glashow resonance events and so-called ``double-bang'' topologies improve the ability to access the flavor of the cosmic high-energy neutrinos and probe the BSM physics. In addition, although details depend on source properties,  Glashow resonance events have the additional feature of being able to inform us of the relative strengths of neutrino and antineutrino emission, which may help us discriminate astrophysical scenarios. 
} 
%We show that the planned upgrade to the IceCube detector--{\it IceCube Gen-2}-- is well-positioned to make dramatic improvements in both directions, thanks to their heightened flavor sensitivity.  Compared to the present incarnation of IceCube, the additional flavor discriminants given by Glashow resonance events and so-called ``double-bang'' topologies significantly improve the ability to access the flavor of the cosmic neutrinos. Glashow resonant events have the additional feature of being able to inform us of the relative strengths of neutrino and antineutrino emission, which was hitherto impossible.  In addition to the possibility of discriminating between $pp$ and $p\gamma$ sources, we find that neutrino decay, Pseudo-Dirac states, and neutrino scattering may be found or strongly constrained at {\it IceCube Gen-2}. 
\end{abstract}
\preprint{}

%\keywords{}

%%%%%%%%%%%%%%%%%%%%%%%%%%%%%%%%%%%%%%%%%%%%%%%%%%%%%%%%%%%%%%%%%%%
\maketitle

%%%%
\section{Introduction}
%%%%
The IceCube collaboration has made the exciting discovery of a new source of ultra-high energy neutrinos~\cite{Aartsen:2013bka,Aartsen:2015ivb,Aartsen:2013jdh,Aartsen:2013eka,Aartsen:2014gkd,Aartsen:2015knd,Aartsen:2015rwa}. 
These neutrinos are of unknown origin, though many possibilities have been suggested~\cite{Murase:2014tsa,Meszaros:2015krr,Waxman:2015ues}. 

A key piece of information contained in these neutrinos is their flavor, which may provide insight into both the source of these neutrinos and to possible new physics in the neutrino sector~\cite{Learned:1994wg,Beacom:2002vi,Beacom:2003eu,Beacom:2003nh,Blum:2007ie,Winter:2012xq,Bustamante:2015waa}. 
For example, astrophysical sources of high-energy neutrinos can produce neutrinos via $pp$ or $p\gamma$ processes. Both $pp$ scenarios~\cite{Murase:2013rfa} and $p\gamma$ scenarios~\cite{Murase:2013ffa,Stecker:2013fxa,Winter:2013cla,Murase:2015xka} have been suggested as viable explanations of the diffuse neutrino flux observed in IceCube.
At sufficiently high energies, $pp$ sources yield a source flavor ratio of $(\alpha_{e}^{S},\alpha_{\mu}^{S},\alpha_{\tau}^{S})=(1,2,0)$ that is symmetric between neutrinos and antineutrinos, thanks to multipion production. 
On the other hand, $p\gamma$ sources may yield $(\alpha_{e}^{S},\alpha_{\mu}^{S},\alpha_{\tau}^{S})=(1,1,0)$ for neutrinos and $(\bar{\alpha}_{e}^{S},\bar{\alpha}_{\mu}^{S},\bar{\alpha}_{\tau}^{S})=(0,1,0)$ for antineutrinos at lowest-order, when the $\Delta$-resonance is dominant.
These differences {may be} large enough to potentially be distinguished in IceCube with future data.  Indeed, present flavor data is already sufficient to strongly disfavor neutron decay source models~\cite{Barger:2014iua,Bustamante:2015waa}. 

Similarly, the flavor content of high-energy neutrinos can probe beyond Standard Model (BSM) processes involving neutrinos. For example, neutrino self-interactions can alter the flavor content of the flux via self-scattering on the cosmic neutrino background~\cite{Ioka:2014kca,Ng:2014pca,Ibe:2014pja,Blum:2014ewa,Cherry:2014xra,Kamada:2015era,DiFranzo:2015qea}.  Other BSM impacts could arise from neutrino decay~\cite{Beacom:2002cb,Baerwald:2012kc}, Pseudo-Dirac neutrinos~\cite{Wolfenstein:1981kw,Petcov:1982ya}, CPT violation~\cite{Barenboim:2003jm,Kostelecky:2003xn,Hooper:2005jp,Anchordoqui:2005gj,Ando:2009ts,Bustamante:2010nq,Arguelles:2015dca}, and BSM modifications of neutrino-nucleus interactions (e.g. leptoquarks)~\cite{Barger:2013pla,Dutta:2015dka,Dey:2015eaa}. {Flavor and spectral analyses have been carried out in a number of works using current IceCube data~\cite{Chatterjee:2013tza,Mena:2014sja,Watanabe:2014qua,Chen:2014gxa,Palladino:2015zua,Palomares-Ruiz:2015mka,Aartsen:2015knd}}.

\begin{figure}[b!]
\begin{center}
 \includegraphics[width=.5\textwidth]{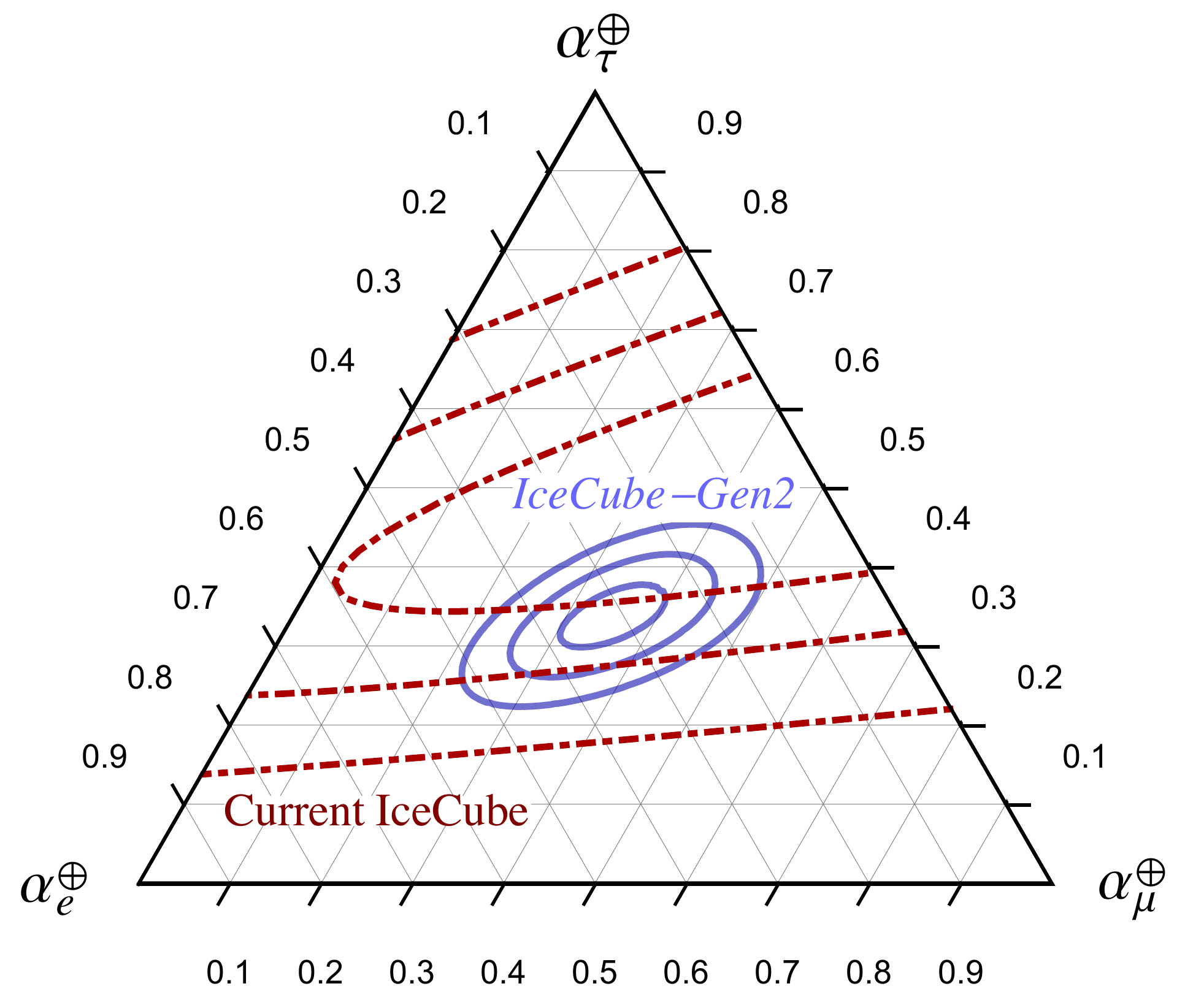} 
\caption{Comparison of present IceCube flavor measurements {based on the combined likelihood analysis~\cite{Aartsen:2015knd}}, and our projected {\it IceCube-Gen2} sensitivity ($(1,2,3)\sigma$ CL contours) based on a high-energy starting event (HESE) analysis described in Sec.~\ref{fit}. 
{We assume equal flavor ratios at the Earth with a $E^{-2.2}$ spectrum above 100 TeV and a 10 year exposure.}}
\label{figintro}
\end{center}
\end{figure}

{In this paper we perform a flavor and spectral analysis for next-generation neutrino detectors such as the {\it IceCube-Gen2} detector, and find striking improvements in BSM physics sensitivity as well as possible source discrimination. {As shown in Fig.~\ref{figintro}, neutrino flavor information can be significantly improved in the {\it IceCube-Gen2} era}.  We show that the flavor sensitivity can be enhanced by Glashow resonance (sensitive solely to $\bar{\nu}_{e}$ via $\bar{\nu}_{e} + e^{-} \rightarrow W^{-}$) and so-called ``double-bang'' events (sensitive to $\nu_{\tau}/\bar{\nu}_{\tau}$ from a high-energy $\tau$ lepton that decays some distance from the production point).} Additional details about our analysis and assumptions can be found in Sec.~\ref{fit}.

The organization of this paper is as follows. In Sec. II, we discuss the variety of source models that can affect flavor ratios. In Sec. III we summarize our mock-up of a {\it IceCube-Gen2}-like detector and statistical analysis. We pay careful attention to the definition and detector sensitivity of four distinct events classes: showers, tracks, so-called double-bangs, and Glashow resonance events. In Sec. IV we present our analysis results from fits the mock data. We demonstrate the added sensitivity afforded by Glashow resonance and double-bang events. In Sec. V we turn to BSM scenarios that can impact the flavor and spectrum of cosmic neutrinos.  Finally we discuss our results and conclude in Secs. VI and VII respectively. 

\section{Flavor Ratio Source Models}
%
%\begin{table}[b]
%\begin{center} 
%    \begin{tabular}{ | l  | p{1.8cm} | p{1.5cm} |}
%    \hline
%       $[N_{GR},N_{DB}]$  & $\Phi \propto E_{\nu}^{-2.2}$ & $\Phi \propto E_{\nu}^{-2.5}$ \\  \hline\hline
%
%    $pp$  & $[23,5]$ & $[8,2]$\\  \hline
%        $pp$ (with $\mu^{+}$ damping)  & $[15,6]$& $[6,2]$  \\ \hline
%    $p \gamma$ (min. $\pi^{-}$) & $[9,5]$ & $[4,2]$\\ \hline
%    neutron decay  & $[73,4]$ & $[28,1]$\\ \hline
%
%       \end{tabular}
%       \caption{Estimated number of Glashow resonance and double-bang events for different choices of terrestrial flavor ratios and neutrino spectrum. In the $p\gamma$ the dominant cross section for neutrino production is from $p\gamma \rightarrow n \pi^{+}$, with the dominant source of $\bar{\nu}_{e}$ originating from $\gamma p \rightarrow p \pi^{+} \pi^{-}$. }
%\label{table}
%\end{center}
%\end{table}

Astrophysical environments have long been considered excellent sources of high-energy neutrinos. The dominant production mechanisms are via hadronuclear mechanisms, so-called $pp$ sources, and photohadronic interactions, or so-called $p\gamma$ sources. 

{For sufficiently high-energy $pp$ interactions}, charged pions are produced in equal quantities and therefore yield source flavor ratios $\alpha_{S}= (1,2,0)$ and identical flavor ratios for antineutrinos, $\bar{\alpha}_{S}= (1,2,0)$. Since neutrinos are produced and detected in flavor eigenstates and the oscillation lengths are very short compared to the travel distance, the terrestrial flavor ratios are incoherently summed
\be \alpha^{i}_{\oplus} = \sum_{j,\mu}  |U_{i\mu}|^{2} |U_{j \mu}|^{2} \alpha^{j}_{S},
\ee
where $U$ is the Pontecorvo-Maki-Nakagawa-Sakata (PMNS) matrix, and $\alpha_{\oplus},\alpha_{S}$ are the terrestrial and source flavor ratios respectively.  We have assumed that wave packets do not overlap and lose coherence, and refer the reader to~\cite{Farzan:2008eg} for a careful discussion of coherence loss for high-energy neutrinos.  The known neutrino mixing parameters approximately coincide with the tribimaximal ansatz: $\delta =0$, $\theta_{13} = 0$, $\theta_{23} = \pi/4$, and $\sin(\theta_{12}) = 1/\sqrt{3}$. With this mixing assumption the source flavor ratios at Earth become, $\alpha_{S}=\bar{\alpha}_{S}= (1/3,2/3,0) \longrightarrow \alpha_{\oplus} = (1/3,1/3,1/3)$, which is the well-known result of equal flavor ratios for $pp$ sources.  Note that abandoning the tribimaximal ansatz and instead using the best-fit mixing angles~\cite{Gonzalez-Garcia:2014bfa} yields $\alpha_{\oplus} = (0.35,0.33,0.32)$, i.e. a $\lesssim 9 \%$ correction to the tribimaximal values.  The effect of the uncertainties on the oscillation parameters has been recently studied in~\cite{Bustamante:2015waa}.

{For $p\gamma$ interactions, details depend on target photon spectra.  If the target photon spectral index $\beta$ is harder than $\beta\sim1$, multipion production becomes relevant (e.g.,~\cite{Murase:2005hy,Baerwald:2010fk}).  In this limit, we expect $\alpha_{S}=\bar{\alpha}_{S}= (1/3,2/3,0) \longrightarrow \alpha_{\oplus} = (1/3,1/3,1/3)$ as in $pp$ scenarios.
For soft target photon spectra}, the dominant cross section for neutrino production is from $p\gamma \rightarrow n \pi^{+}$, with the dominant source of $\bar{\nu}_{e}$ originating from $p\gamma\rightarrow p \pi^{+} \pi^{-}$.  
{Assuming sufficiently soft target photon spectra, we consider} the contributions from both $\sigma(p \gamma \rightarrow p \pi^{+} \pi^{-})$ and $\sigma(p\gamma \rightarrow n \pi^{+})$ in our estimate of a realistic $p\gamma$ source.  As is well-known the dominant contribution is $\pi^{+} \rightarrow \mu^{+} \nu_{\mu} \rightarrow e^{+} \nu_{e} \nu_{\mu} \bar{ \nu}_{\mu}$. This yields $\alpha^{(0)}_{S} = (1,1,0)$ and $\bar{\alpha}^{(0)}_{S} = (0,1,0)$. The sub-dominant charged-pion pair production instead yields equal amounts of $e^{+} \nu_{e} \nu_{\mu} \bar{ \nu}_{\mu}$ and $e^{-} \bar{\nu}_{e} \nu_{\mu} \bar{ \nu}_{\mu}$, or in other words: $\alpha_{S}^{(1)} = (1,2,0)$ and $\bar{\alpha}_{S}^{(1)} = (1,2,0)$. See the Appendix for additional details on the charged pion fractions. 
%\com{(I am confused. Should we have $\alpha_{S}^{(1)} = (0,1,0)$ and $\bar{\alpha}_{S}^{(1)} = (0,1,0)$? Then we have $\alpha_{S}^{(0)}+\alpha_{S}^{(1)} = (1,2,0)$ in the limit of ${\mathcal N}_+={\mathcal N}_-$.)}

Flavor ratios can be further modified by astrophysical effects, including cooling of pions and muons~\cite{Kashti:2005qa,Ando:2005xi}, kaon and charm meson contributions~\cite{Asano:2006zzb,Enberg:2008te}, and reacceleration of pions and muons~\cite{Murase:2011cx,Winter:2014tta,Kawanaka:2015qza}.  In some specific models such as the choked jet model~\cite{Meszaros:2001ms,Murase:2013ffa}, even matter effects in the neutrino oscillation can affect the resulting flavor ratio~\cite{Razzaque:2009kq}. 
This work focuses on probing BSM neutrino physics rather than these astrophysical effects, so considering the simplest $pp$ and $p\gamma$ scenarios is enough for the purpose of the present work.

\section{Fit Setup and Experimental Details}
\label{fit}
The main advantage of the {\it IceCube-Gen2 } detector is in the increase in effective areas and volumes, {and the improvement in its angular resolution}. This gives the detector much greater sensitivity to new flavor probes like Glashow resonance and double-bang events.  However for double-bang events, the increase in string spacing reduces sensitivity somewhat since the energy threshold is determined by the requirement that the produced $\tau$ travels at least the distance between two adjacent strings. 

{The effective volume is increased by a factor of 7.7 and 11.3 for 240m and 300m spacing, respectively. 
Correspondingly, the effective areas are increased by a factor of 4 and 5, respectively over the 86-string configuration of IceCube (IC-86)~\cite{Aartsen:2014njl}.} Then for our mock-up of the {\it IceCube-Gen2 } we simply scale up the effective areas of IC-86~\cite{Aartsen:2013jdh}.  Using this information with the mock-up procedure below, we obtain Glashow resonance event rates in agreement with the collaboration's Table I in~\cite{Aartsen:2014njl}  at the $(12-18)\%$ level depending on the spectral tilt with better agreement for steeper spectra, but with agreement at the $(8-9)\%$ level for IC-86. 

We mock up the number of different classes of events at IceCube in the following simplified manner. We will fit to the four main observables of flavor at {\it IceCube-Gen2}: showers, tracks, Glashow resonance events~\cite{Anchordoqui:2004eb,Bhattacharya:2011qu,Barger:2014iua}, and double-bang topologies~\cite{Learned:1994wg,Beacom:2003nh,Bugaev:2003sw,Jones:2003zy}.

It is important to highlight two subtleties that we take into account in our setup: (1) we account for the fact that the neutrino energy does not in general coincide with the energy deposited in the detector and the deposited energy depends on the nature of the interaction; and (2) the neutrinos are opaque to the earth in an energy-dependent and flavor-dependent manner. We account for terrestrial opacity by using the effective areas provided by the IceCube collaboration~\cite{Aartsen:2013jdh} which also include information on the neutrino-nucleus, neutrino-electron interactions, as well as selection cuts of the high-energy starting event (HESE) analysis. 
We adopt this ``effective area'' method for the main body of paper, but show more optimistic results in~\ref{sec:discuss}. 

To properly account for the difference between neutrino energy and the energy deposited in the detector, we must consider the different ways in which each flavor can interact.  
First, consider the so-called ``shower events'' which are a combination of 
\be
N_{S} = N_{S}^{{\rm NC,all}} + N_{S}^{{\rm CC,e}} + N_{S}^{{\rm CC,\tau_{had}}} + N_{S}^{{\rm CC,\tau_{\ell}}}
\ee 
where $N_{S}^{{\rm CC,e}}$ is the number of $\nu_{e}+\bar{\nu}_{e}$ induced CC showers, $N_{S}^{{\rm CC,\tau_{had}}}$ is the number of $\nu_{\tau}+\bar{\nu}_{\tau}$ induced CC showers with hadronic $\tau$ decays, $N_{S}^{{\rm CC,\tau_{\ell}}}$ is the number of $\nu_{\tau}$ induced CC showers with leptonic $\tau$ decays, and $N_{S}^{{\rm NC,all}}$ are the number of showers from NC interactions of any neutrino flavor. 

Following Ref.~\cite{Blum:2014ewa}, we calculate shower and track event rates. For shower events, we estimate these sub-contributions as 

%\begin{widetext}
%
%\[

\onecolumngrid

\begin{align}
N_{S}^{{\rm NC,all}} (E_{1},E_{2}) &= p_{NC} \sum_{i = e,\mu, \tau} 4 \pi T \int_{E_{1}/\beta_{NC}}^{E_{2}/\beta_{NC}} dE~ ( \alpha_\oplus^{i} + \bar{\alpha}_\oplus^{i} )\Phi_{\nu}(E) A_{i}(E) \\
N_{S}^{{\rm CC,\tau_{had}}} (E_{1},E_{2}) &= \left(1-p_{NC}\right)p_{had} 4 \pi T \int_{E_{1}/\beta_{had}}^{E_{2}/\beta_{had}} dE~ ( \alpha_\oplus^{\tau} + \bar{\alpha}_\oplus^{\tau} ) \Phi_{\nu}(E) A_{\tau}(E)  \\
N_{S}^{{\rm CC,\tau_{lep}}} (E_{1},E_{2}) &= \left(1-p_{NC}\right)(1-p_{had}) 4 \pi T \int_{E_{1}/\beta_{lep}}^{E_{2}/\beta_{lep}} dE~ ( \alpha_\oplus^{\tau} + \bar{\alpha}_\oplus^{\tau} ) \Phi_{\nu}(E) A_{\tau}(E) \\
 N_{S}^{{\rm CC,e}} (E_{1},E_{2}) &= \left(1-p_{NC}\right)  4 \pi T \int_{E_{1}}^{E_{2}} dE~ ( \alpha_\oplus^{e} + \bar{\alpha}_\oplus^{e} ) \Phi_{\nu}(E) A_{e}(E) 
\end{align}
%
%\]
%\end{widetext}

\twocolumngrid
{where $A_{i}(E)$ is the effective area of neutrino flavor $i$, $T$ is the exposure time,} $p_{NC} = \sigma_{NC}/(\sigma_{CC}+\sigma_{NC}) \simeq 0.28$, $p_{had} \simeq 0.65$, $ \beta_{NC} = \langle y \rangle E_{\nu}$, $ \beta_{had} =\left[\langle y \rangle + \frac{2}{3}(1-\langle y \rangle)\right]  E_{\nu}$, and $ \beta_{lep} =\left[\langle y \rangle + \frac{1}{3}(1-\langle y \rangle)\right]  E_{\nu}$~\cite{Gandhi:1995tf,Gandhi:1998ri}. Note that $\Phi_{\nu}(E)$ is the all-flavor flux of neutrinos and anti-neutrinos. We use the energy-dependent mean inelasticities from~\cite{Gandhi:1995tf} for neutrinos and anti-neutrinos. Notice that charged-current electron neutrino initiated cascades deposit all of the available neutrino energy. The shower channel is very important given the low atmospheric backgrounds, as considered in~\cite{Beacom:2004jb}.

Similarly, $\nu_{\mu}$ CC interactions yield striking track events. These are estimated as 
\be 
N_{T} = 4 \pi T  \int_{E_{1}/\beta_{\mu}}^{E_{2}/\beta_{\mu}} dE~ (1-p_{NC}) ~( \alpha_\oplus^{\mu} + \bar{\alpha}_\oplus^{\mu} )~\Phi_{\nu}(E)~ A_{\mu}(E) 
\ee
where $ \beta_{\mu} = \langle y \rangle E_{\nu}$.

The Glashow resonance, $\bar{\nu}_{e} + e^{-} \longrightarrow W^{-}$, gives a large enhancement in the detectability of the $\bar{\nu}_{e}$ flux. The number of Glashow resonance events is estimated as
\be   
 N_{GR} = 4 \pi T \int_{E_{G,-}}^{E_{G,+}} dE~ ~\bar{\alpha}_{\oplus}^{e}~\Phi_{\nu}(E)~ A_{e}(E) 
 \ee
 where we take $E_{G,\pm} = E_{G} \pm \Delta E$, where $\Delta E =1$ PeV~\cite{Aartsen:2013jdh} and the resonance energy is $E_{G} = \frac{m_{W}^{2}}{2m_{e}} \simeq 6.3~{\rm PeV}$. 
 
The final class of events we consider are those of the so-called ``double-bang'' topology in which an incoming $\nu_{\tau}$ produces a cascade through a CC interaction. The on-shell $\tau$ lepton then travels a resolvable distance away from the first cascade before decaying hadronically and producing a second cascade. We estimate the number of these double-bang events via
 \be   
 N_{DB} = 4 \pi T \int  dE~ ~( \alpha_\oplus^{\tau} + \bar{\alpha}_\oplus^{\tau} )~\Phi_{\nu}(E)~ A_{\tau}(E) \Theta(E-E_{th})
 \ee
 where the unit-step function accounts for the fact the $\tau$ lepton must be sufficiently energetic for each cascade to be separately resolved. This will be the case for $\tau$ leptons which have decay lengths longer than the string separation in the detector. We therefore adopt, $E_{th} = \lambda_{st}\Gamma_{\tau} m_{\tau}$ where $\lambda_{st}$ is the string spacing, and $m_{\tau}$, $\Gamma_{\tau}$ are the mass and decay width of the tau lepton respectively. For the fiducial 240 m string spacing we consider here this implies a 5.3 PeV threshold. Note that the flavor sensitivity for $\tau$ neutrinos that we adopt here should be viewed as very conservative. 
{Not only double-bang events but also lollipop, inverted-lollipop, and sugardaddy events should be searched for. Also, ``double pulse'' waveforms may allow for $\tau$ discrimination down to $\mathcal{O}(100)$ TeV~\cite{Aartsen:2015dlt}.} 

\begin{figure*}[t]
\begin{center}
 \includegraphics[width=.32\textwidth]{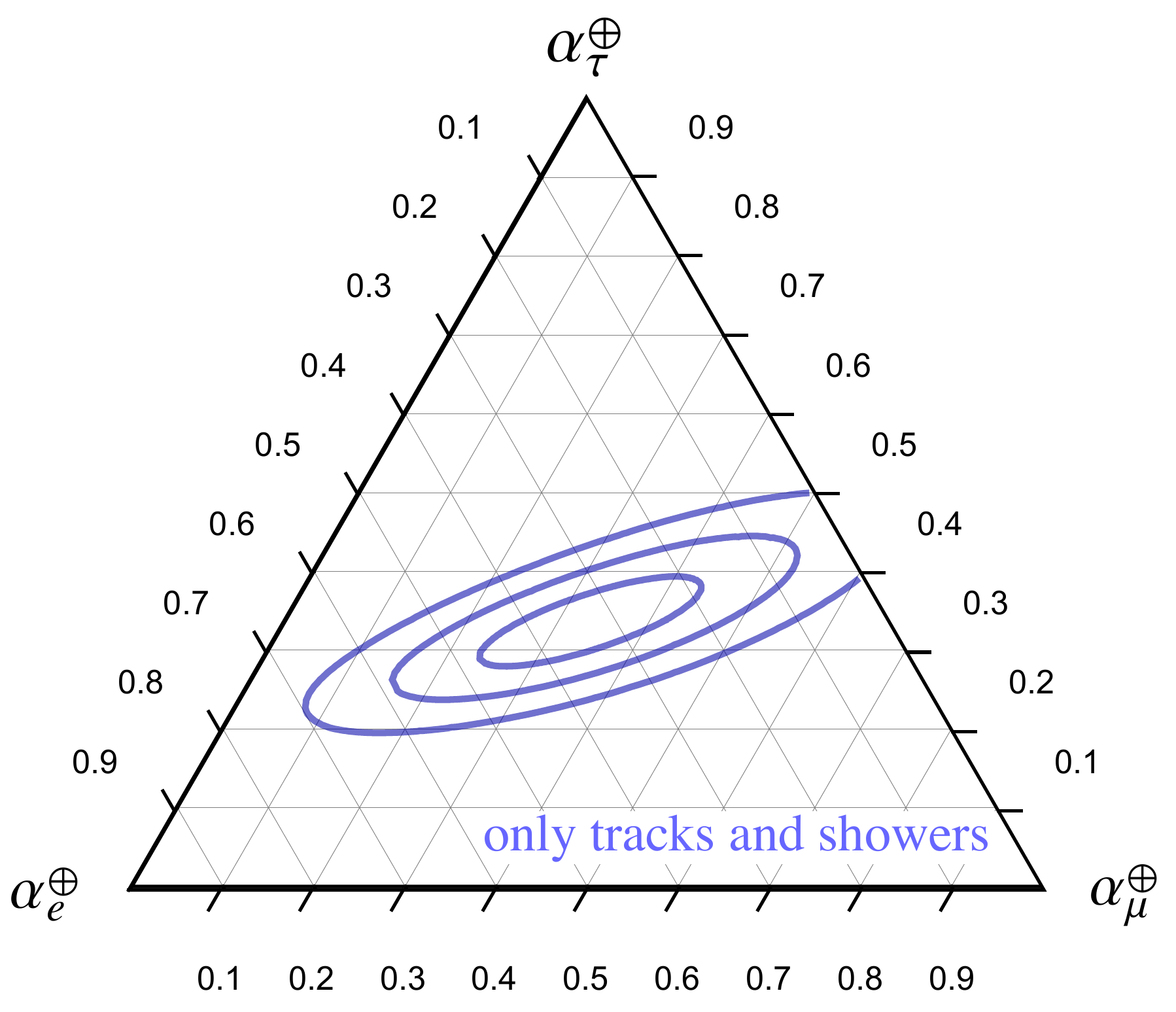} 
  \includegraphics[width=.32\textwidth]{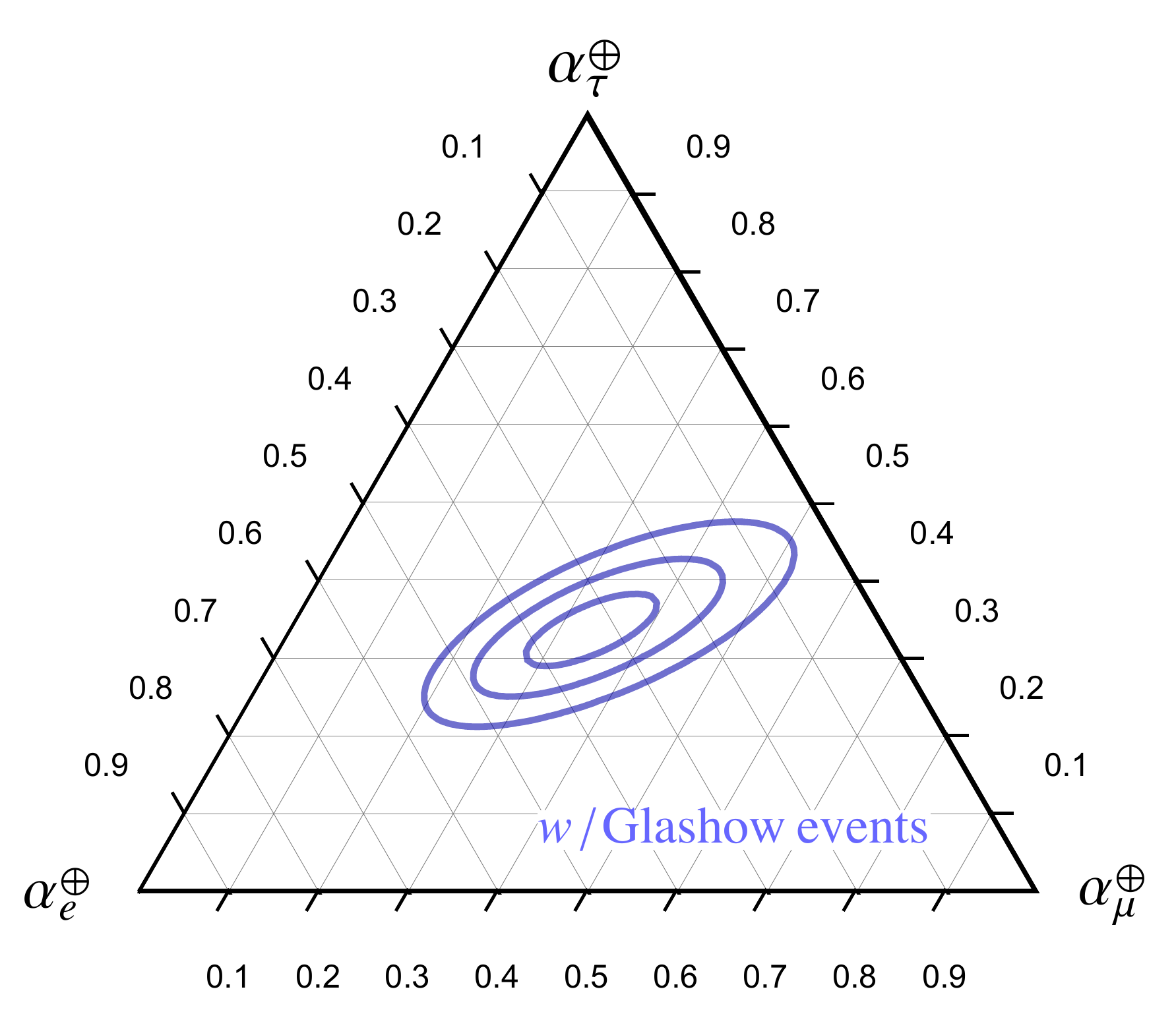} 
 \includegraphics[width=.32\textwidth]{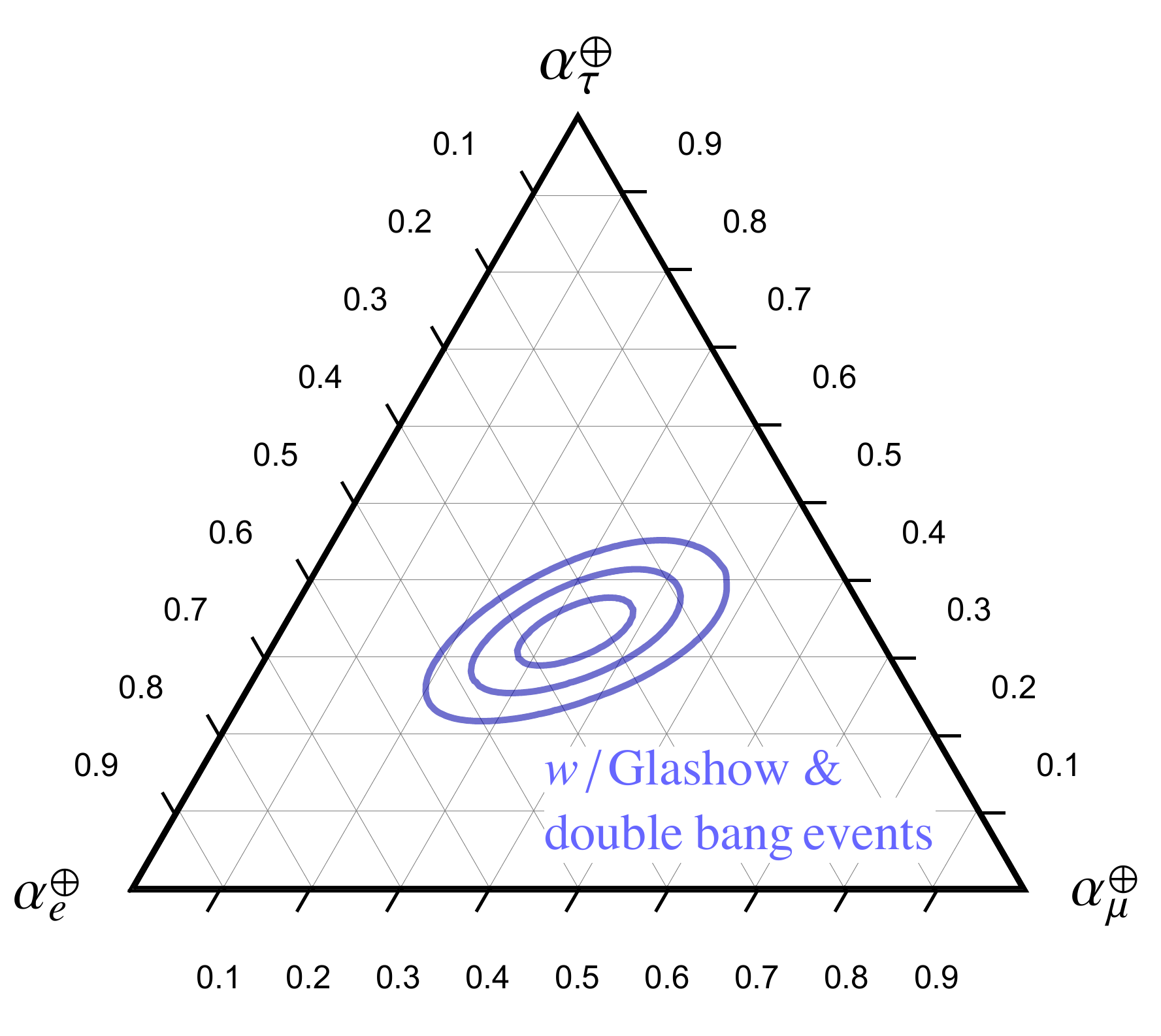} 
  \includegraphics[width=.32\textwidth]{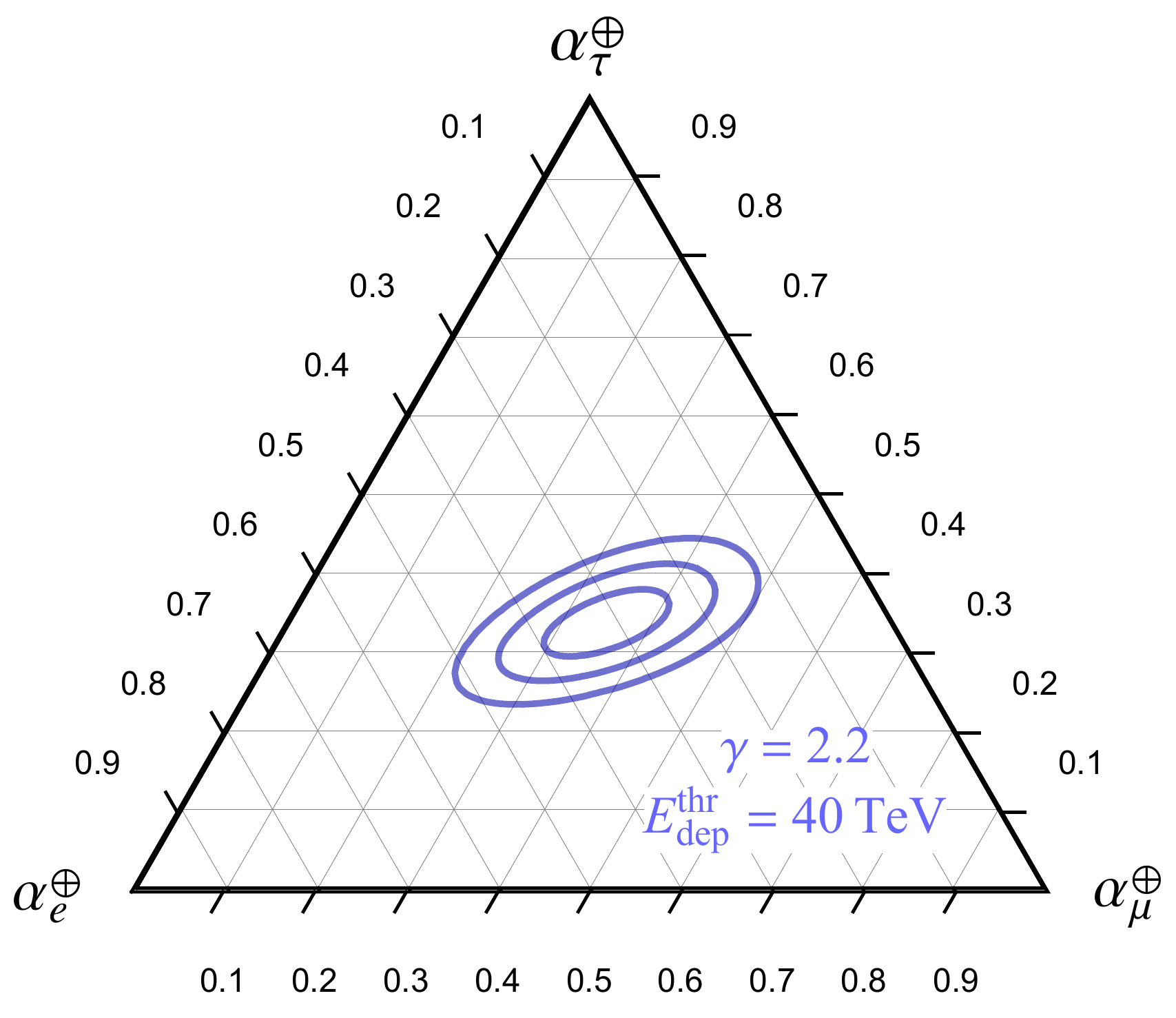} 
  \includegraphics[width=.32\textwidth]{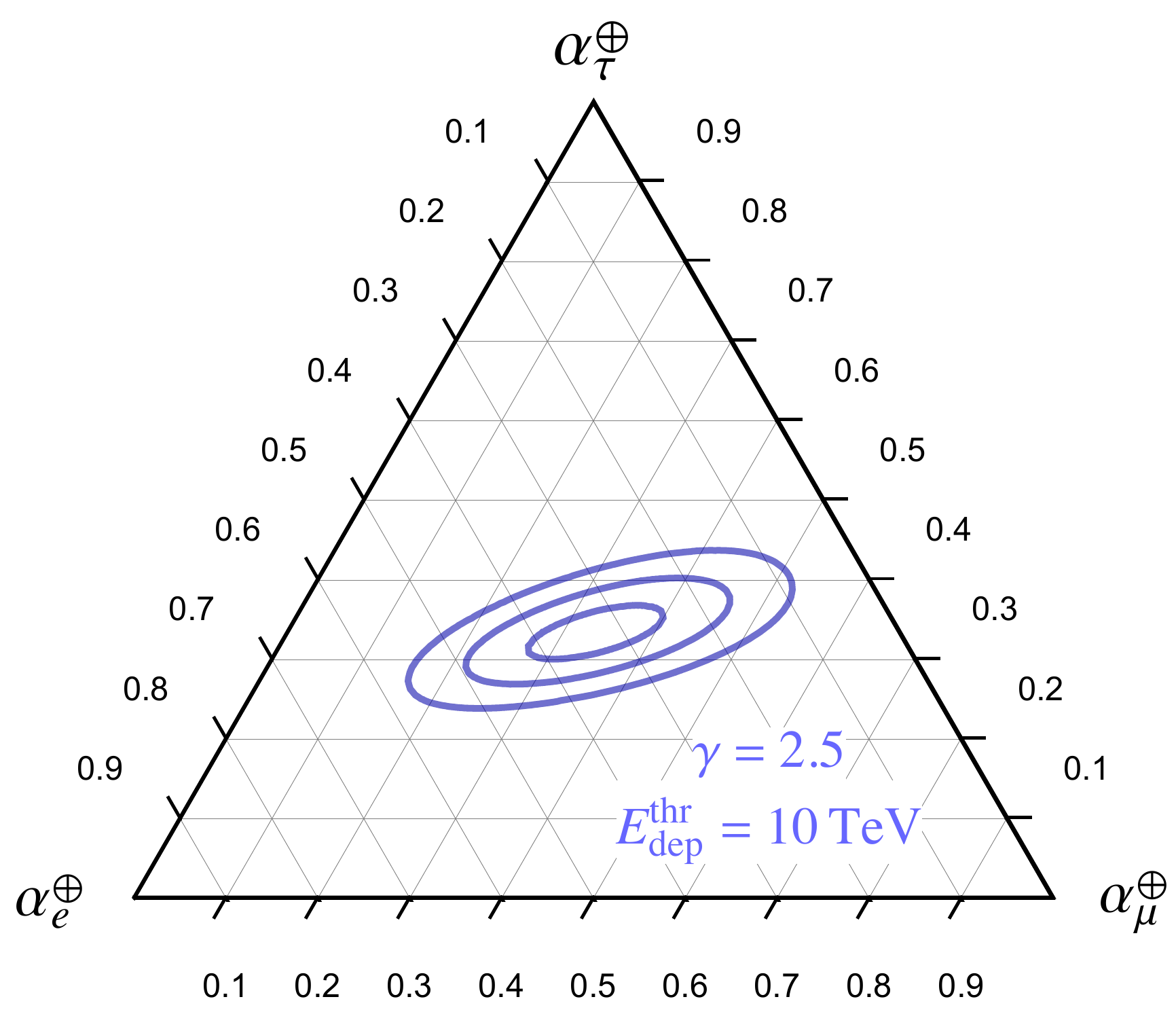} 
  \includegraphics[width=.32\textwidth]{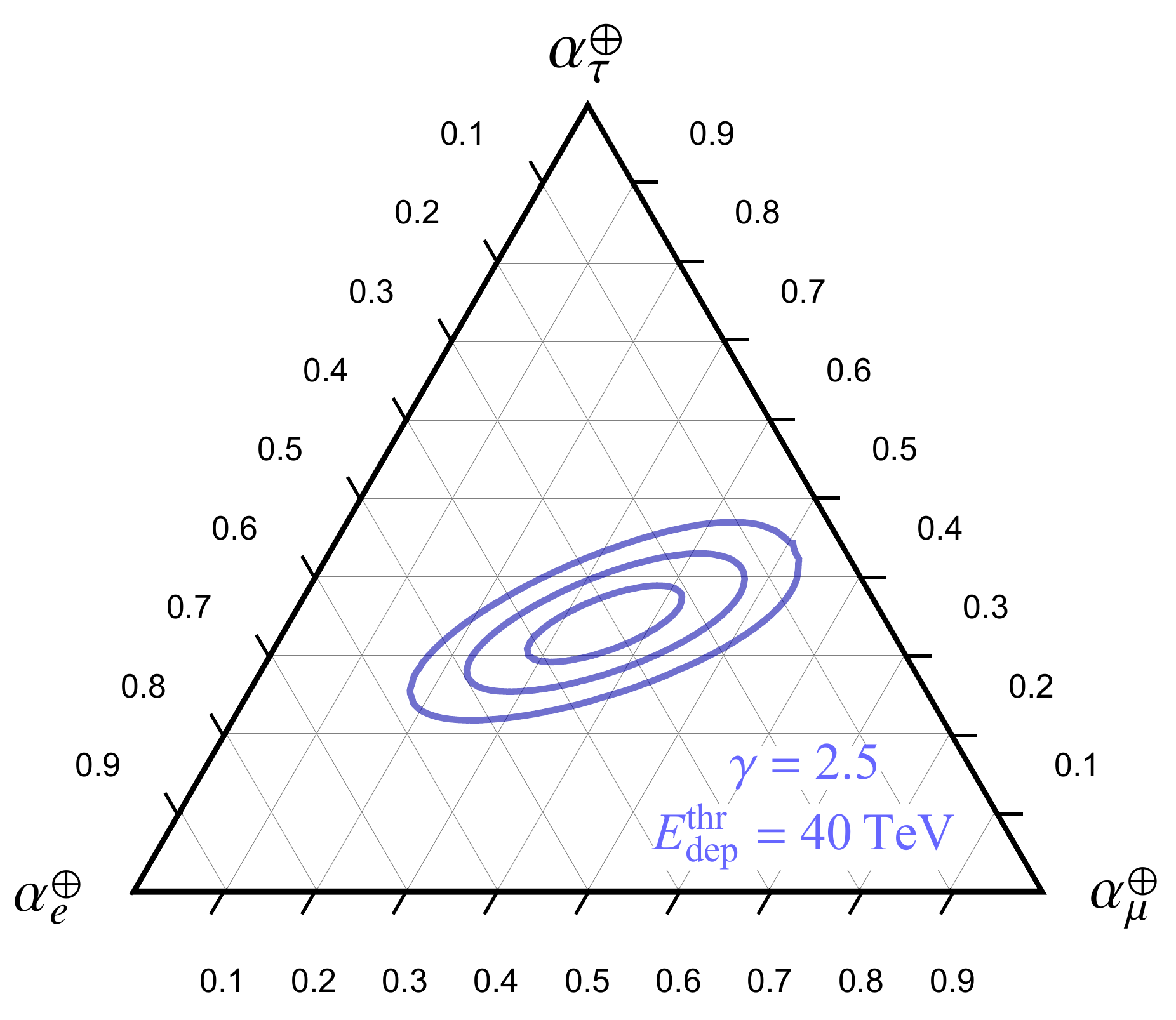} 
\\
\caption{Analysis results displaying the $1\sigma, 2\sigma, 3\sigma$ CL regions using total number of tracks and showers (left) as flavor discriminants, using Glashow resonance events (center), and including double-bang events (right) with 10 years of data at {\it IceCube-Gen2}. Here the input data was produced from a spectrum with $\gamma =2.2$ and equal flavor ratios in the energy range $E_{{\rm dep}} = [100~{\rm TeV}, 10~{\rm PeV}]$.}
\label{fig111}
\end{center}
\end{figure*}

\begin{figure}[b]
\begin{center}
 \includegraphics[width=.4\textwidth]{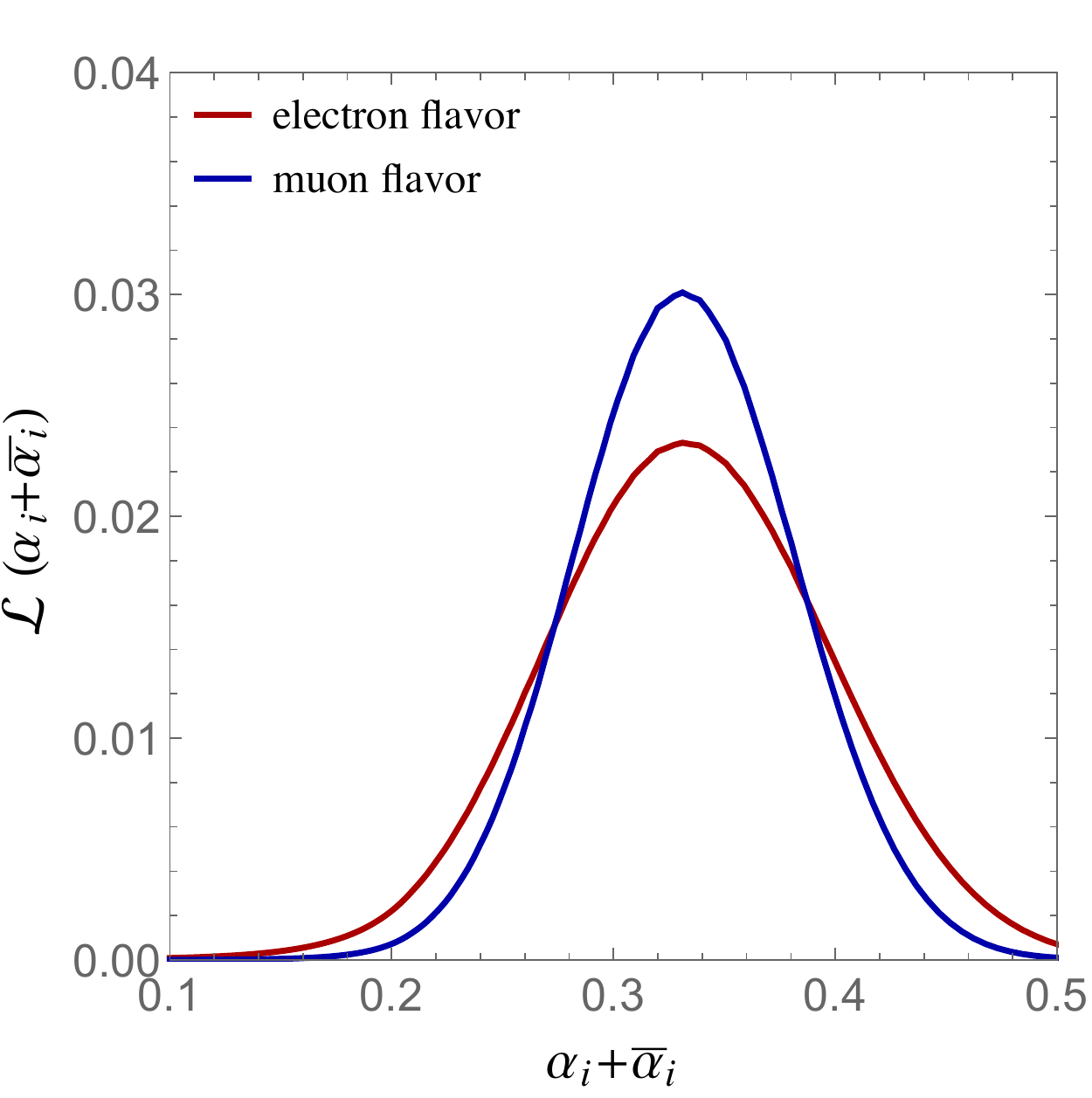} 
\caption{Marginalized likelihood functions of $e$ and $\mu$ flavor ratios for the example shown in the right panel of Fig.~\ref{fig111}. This is a $pp$ source with a $\gamma=2.2$ spectrum where we have included all flavor discriminants in the fit.}
\label{figLike}
\end{center}
\end{figure}

We then use mock data to estimate the statistical likelihood that a given number of tracks, showers, Glashow resonance resonance, and double-bang events have come from a given set of flavor ratios as,
\be \mathscr{L}(\alpha_{e},\alpha_{\mu},\alpha_{\tau}) = \prod_{i}  \mathscr{L}_{i} (\alpha_{e},\alpha_{\mu},\alpha_{\tau})
\ee
where the index $i$ runs over the event classes: showers, tracks, Glashow events, and double-bangs. The likelihood of a given event class is modelled a Poisson process
\be 
\mathscr{L}_{i} (\alpha_{e},\alpha_{\mu},\alpha_{\tau}) = N_{i}(\alpha_{e},\alpha_{\mu},\alpha_{\tau})^{n_{i,true}}\times \frac{\exp^{- N_{i}(\alpha_{e},\alpha_{\mu},\alpha_{\tau})}}{n_{i,true}!}
\ee
where $n_{i,true}$ represent the input number of events in class $i$ while $N_{i}(\alpha_{e},\alpha_{\mu},\alpha_{\tau})$ denotes the number of events in the fit from a given choice of flavor ratios. 

Of course the total event rate is a sum of signal and background events. The background event rates are calculated in a fashion similar to the signal events. In particular we include here the atmospheric neutrino backgrounds~\cite{Honda:2006qj,Enberg:2008te,Abbasi:2011jx} but for simplicity ignore the muon backgrounds which are important for lower energy thresholds. This data is then binned into 5 logarithmically spaced bins per energy decade, and each binned likelihood can be multiplied together to form the final likelihood. 

Lastly, we note that IceCube sometimes mis-reconstructs muon tracks as showers. This has been demonstrated to play an important role in the current IceCube flavor reconstruction~\cite{Palomares-Ruiz:2015mka}. Moreover, occasionally $\tau$ charged current events contribute to track events. 

\begin{figure}
\begin{center}
 \includegraphics[width=.40\textwidth]{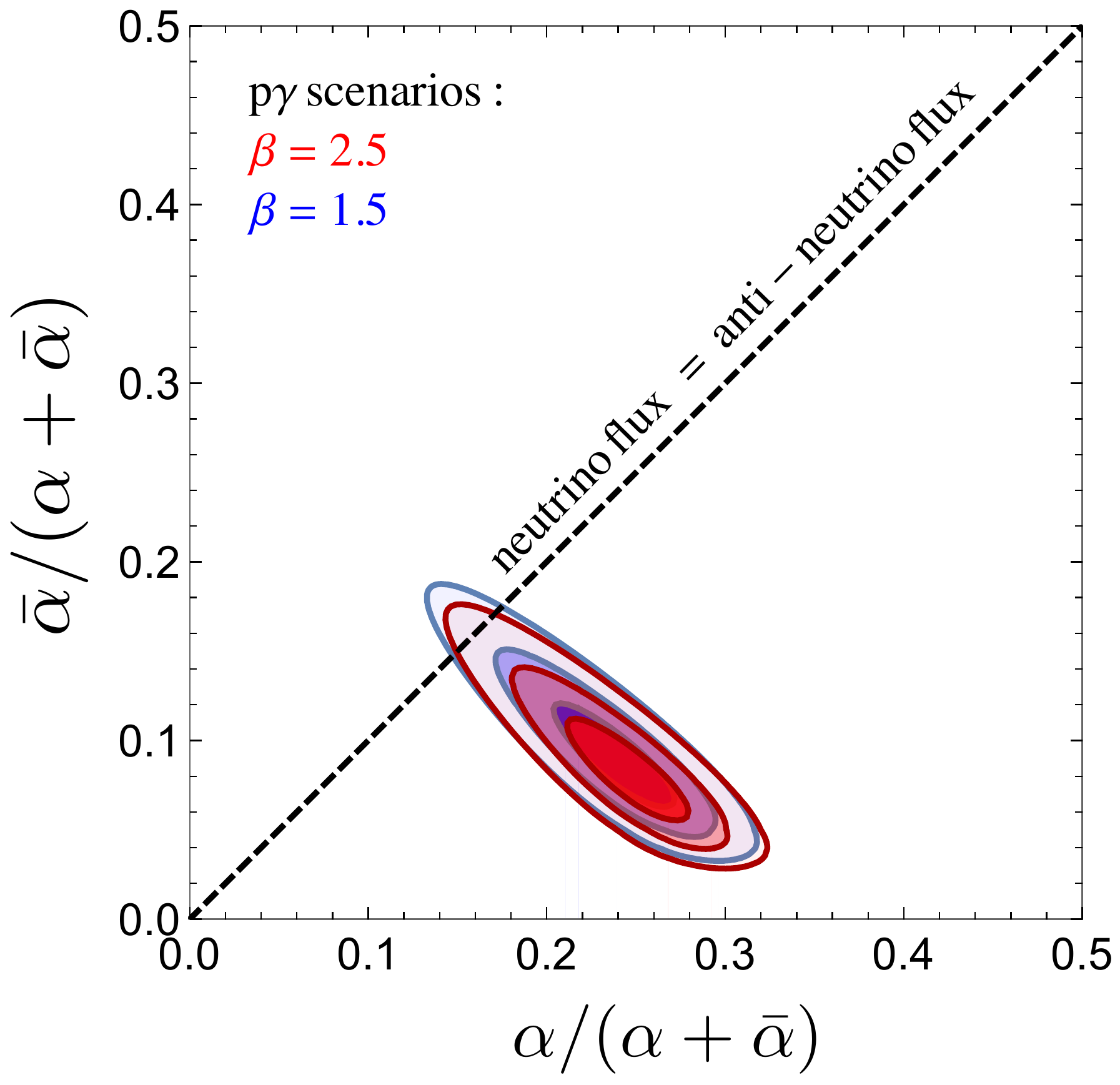} ~~~
\caption{{Flavor reconstruction in $p \gamma$ scenarios.} The canonical $\beta =2$ case interpolates between these contours. }
\label{pgamma}
\end{center}
\end{figure}

\section{Analysis Results}
In addition to uncertainties in the flavor composition of the neutrinos, their spectrum also contains significant uncertainties at present. Let us review the present state of what is known about the spectrum of cosmic neutrinos that IceCube has found.

Typically, the spectral fit to data is reported in terms of an unbroken power law,
\be 
\Phi_{\nu}(E) = \Phi_{0} \left(\frac{100~{\rm TeV}}{E}\right)^{\gamma}.
\ee
The precise values of the spectral index $\gamma$ and normalization $\Phi_{0}$ are not precisely known yet. For example, the central value of the best-fit from the combined likelihood analysis of Ref.~\cite{Aartsen:2015knd}, which found $\phi_{0} = \left(6.7^{+1.1}_{-1.2}\right)\times 10^{-18}~{\rm GeV}^{-1}{\rm s}^{-1} {\rm sr}^{-1}{\rm cm}^{-2}$ and spectral index $\gamma = 2.50 \pm 0.09$. On the other hand, the muon neutrino data sample using northern hemisphere data prefers a more shallow spectral index~\cite{Aartsen:2015rwa}: $\phi_{0} = 5.1^{+1.8}_{-1.8} \times 10^{-18}~{\rm GeV}^{-1}~{\rm cm}^{-2}~{\rm s}^{-1}~{\rm sr}^{-1}$, and $\gamma = 2.2 \pm 0.2$. We shall consider the spectral index to vary in this window, $\gamma =2.2 - 2.5$. Of course, additional spectral features such as a break or exponential cutoff are also possible~\cite{Laha:2013eev}. These possibilities are not yet necessary with present data and will be further clarified as more data accumulates. 
Note that although preliminary, new IceCube data may allow for yet harder spectra above 100 TeV~\cite{Aartsen:2015zva}. 

Finally, notice that in contrast with~\cite{Aartsen:2015knd} we only allow flavor variables to vary in our likelihood fits and do not simultaneously perform a spectral fit. We leave an analysis similar to ~\cite{Aartsen:2015knd} for future work.

%%%%%
\subsection{Astrophysical Effects}
%%%%%

\begin{table}[b]
\begin{center} 
    \begin{tabular}{ | l  | p{1.8cm} | p{1.5cm} |}
    \hline
       $[N_{GR},N_{DB}]$  & $\Phi \propto E^{-2.2}$ & $\Phi \propto E^{-2.5}$ \\  \hline\hline

    $pp$  & $[23,5]$ & $[8,2]$\\  \hline
        $pp$ (with $\mu^{+}$ damping)  & $[15,6]$& $[6,2]$  \\ \hline
    $p \gamma$ ({\rm canonical} $\pi^{-}$) & ${[11,6]}$ & {$[4,2]$}\\ \hline
    neutron decay  & $[73,4]$ & $[28,1]$\\ \hline

       \end{tabular}
       \caption{Estimated number of Glashow resonance and double-bang events for different choices of terrestrial flavor ratios and neutrino spectrum at {\it IceCube-Gen2} with a 10 year exposure. In the $p\gamma$ the dominant cross section for neutrino production is from $p\gamma \rightarrow n \pi^{+}$, with the dominant source of $\bar{\nu}_{e}$ originating from $p\gamma \rightarrow p \pi^{+} \pi^{-}$. }
\label{table}
\end{center}
\end{table}

An important question to the BSM sensitivity of neutrino telescopes is the extent to which our results depend on the details of the astrophysical sources (i.e. $pp$ or $p\gamma$ scenarios). Moreover, the same flavor information that we will use to search for BSM physics can also be used to gain insight into the nature of the source. For previous work on distinguishing between $pp$ and $p\gamma$ scenarios (see e.g.~\cite{Winter:2012xq}).

In Table~\ref{table} we summarize the expected Glashow resonance and double bang events for a variety of source models.  The first important observation is that the spectral index will of course play a very crucial role in our ability to determine cosmic neutrino flavor ratios. For example, with $\gamma =2.5$ the distinction between $pp$ and $p\gamma$ sources will be challenging. The second important point from Table~\ref{table} is that with the more optimistic spectral index $\gamma =2.2$ the discrimination between all the source models is significantly improved. Most of this discrimination power is due to the Glashow resonance events. 

Next, let us attempt a more detailed determination of flavor from mock data. As a first illustration we follow the existing literature in fitting to the neutrino and anti-neutrino summed flavor ratios, $(\alpha_{e}+\bar{\alpha}_{e},\alpha_{\mu}+\bar{\alpha}_{\mu},\alpha_{\tau}+\bar{\alpha}_{\tau})$. This is a reasonable procedure in the case of $pp$ sources. This reduces the six flavor ratios to effectively only two independent combinations (since they sum to unity, i.e. $\alpha_{\tau}+\bar{\alpha}_{\tau} = 1- (\alpha_{e}+\bar{\alpha}_{e}+\alpha_{\mu}+\bar{\alpha}_{\mu})$ is redundant). We display the resulting fits under this assumption in Fig.~\ref{fig111} with mock data generated under the assumption of a $pp$ source without $\mu$ damping. 

First, let us focus on the top panel of Fig.~\ref{fig111}. Here we have fixed the spectral index to $\gamma =2.2$ and kept the analysis energy region to be $[100~{\rm TeV}, 10^{4}~{\rm TeV}]$.  In going from the left (tracks and showers only), to center (tracks, showers, and Glashow events), to right panels (tracks, showers, Glashow, and double bang events) we illustrate the increased flavor sensitivity afforded by the Glashow and double-bang events. In the remainder of the paper we will therefore include all event types in our flavor fits.

Next, we would like to examine the effect of varying the above assumptions by considering a different spectral index and energy threshold. This is illustrated in the bottom panel of Fig.~\ref{fig111}. In the left figure, we keep $\gamma =2.2$ and show the new allowed regions with a lowered energy threshold of 10 TeV. Although this may be overly optimistic, it gives a sense of the possible improvement achieved with a lowered energy threshold. This should be compared against the right figure in the top panel of Fig.~\ref{fig111} which has a 100 TeV threshold. We display the analogous comparison for an index $\gamma =2.5$ in the center and right figures in the bottom panel which has an energy threshold of 10 and 100 TeV respectively. 

We can also see what can be determined about an individual flavor ratios by marginalizing over the other two. We do this for the $\gamma =2.2$ example (top right figure in Fig. ~\ref{fig111}), and display the resulting marginalized likelihood functions in Fig.~\ref{figLike}. This yields 
 $\alpha_{e}+\bar{\alpha}_{e} =0.33 \pm 0.03$ and $\alpha_{\mu}+\bar{\alpha}_{\mu} =0.33 \pm 0.04$ (1$\sigma$ errors). We therefore obtain information very consistent with a $pp$ source. Recall that this result was obtained with a spectral index, $\gamma = 2.2$. The analogous exercise for the $\gamma = 2.5$ case results in $\alpha_{e}+\bar{\alpha}_{e} =0.33 \pm 0.06$, $\alpha_{\mu}+\bar{\alpha}_{\mu} =0.33 \pm 0.05$.

\begin{figure*}[t]
\begin{center}
 \includegraphics[width=.4\textwidth]{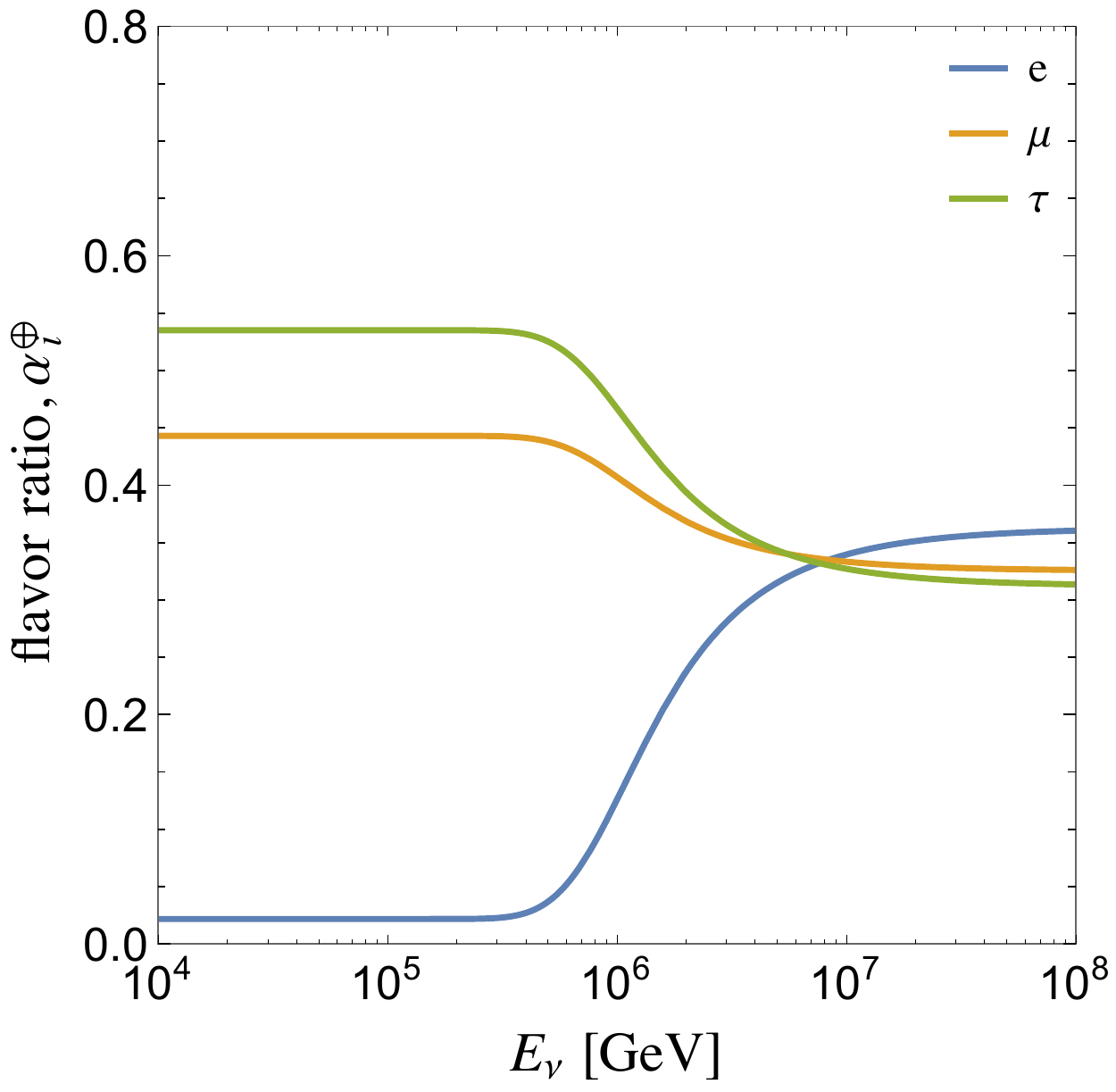} ~~~
 \includegraphics[width=.45\textwidth]{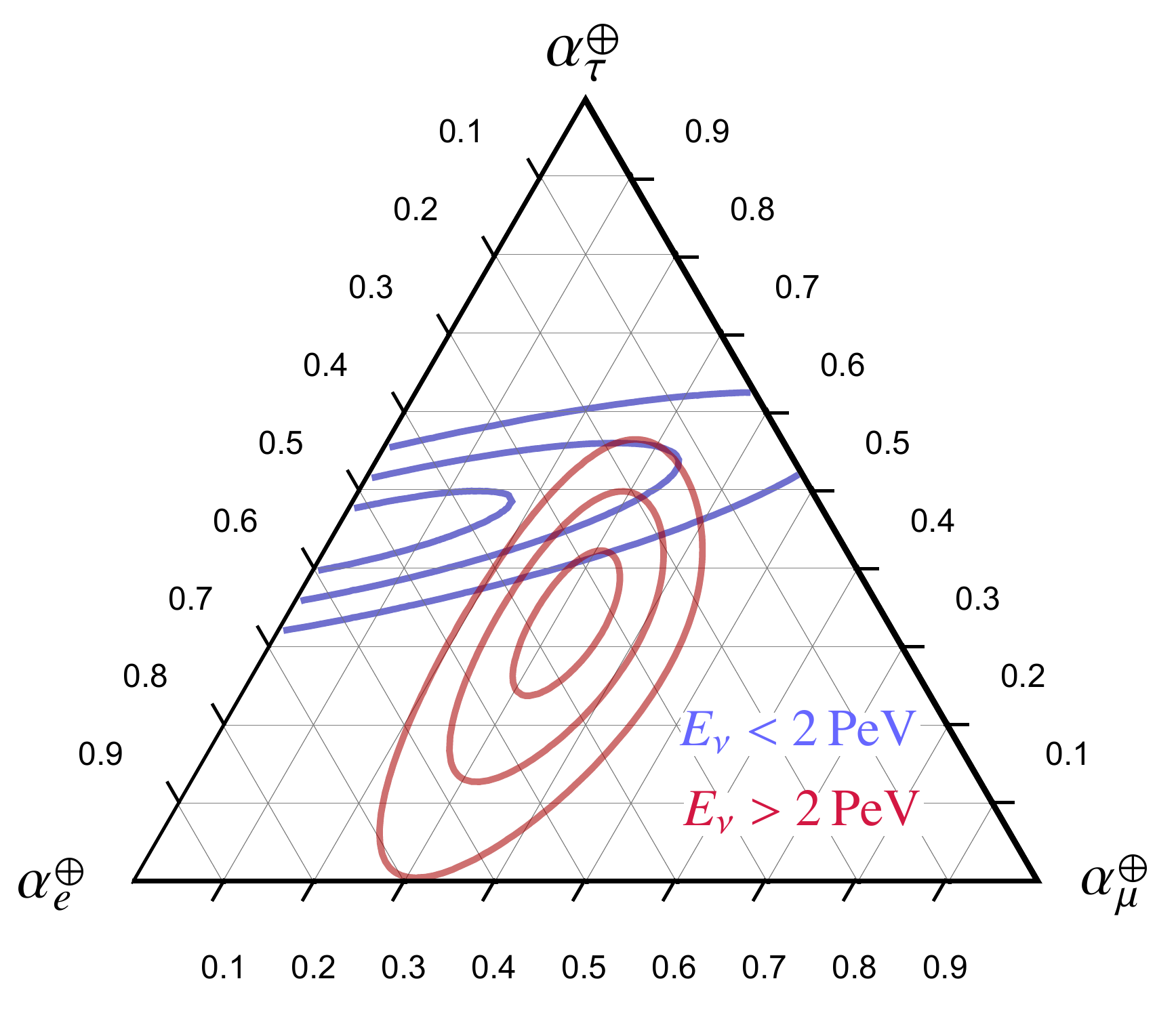} ~~~

\caption{{\it Left panel:} Here we display an illustrative example of incomplete neutrino decay in which $\nu_{1}$ and $\nu_{2}$ decay in the IH. The source has been chosen to produce $1:2:0$ flavor ratios. {\it Right Panel}: Here we show the projected {\it IceCube-Gen2} sensitivity. We have imposed $\sum_{i}\alpha_{i} = 1$, in the left panel, but note that neutrino decays induce an overall flux suppression on low energies since it is only the $\nu_{3}$ state that exists at low energies whereas the other two are present for higher energies. %May be problems with total flux limit \com{??: Note that the best-fit spectral index with present data in such a case is steeper than the no-decay case since the total flux rises at $\gtrsim {\rm PeV}$ energies. }
{In this example we have fixed $\kappa^{-1} = 10^{2}~{\rm s}/{\rm eV}$.} 
{The star-formation rate is used as a redshift evolution of the sources.}
}
\label{nudecay}
\end{center}
\end{figure*}

Next we investigate what {\it IceCube-Gen2} can say about $p\gamma$ sources at the level of a detailed flavor fit. The above fitting procedure is not well-suited to a $p\gamma$ source since it assumes that flavor ratios are the same for neutrinos and antineutrinos. 
As a canonical example, we take the photon spectral index to be $\beta=2$.  In contrast with $pp$ scenarios, a better characterization of the $p\gamma$ case is that all flavor ratios for neutrinos are equal to one common value, $\alpha \simeq 0.26$, while the flavor ratios of antineutrinos equal to another common value, $\bar{\alpha} \simeq 0.08$. This is illustrated in Fig.~\ref{pgamma}. We see that a $\sim 3~\sigma$ deviation from the $pp$ expectation of equal neutrino and antineutrino fluxes is observed. However as can be seen from the figure, not enough in the flavor ratios to have sensitivity to the photon index.

%\com{The following paragraph could go to Appendix with the bottom panel of Fig.4}
%Finally, we entertain the possibility that the relative abundance of $\pi^{-}$ and $\pi^{+}$ could be determine from flavor information for a p$\gamma$ source (see Eq.~\ref{eq:pgamma}). The resulting likelihood functions are shown in Fig.~\ref{pgamma} for a variety of input pion ratios. We see that the none of the likelihoods are very peaked, and that no strong conclusion can be reached for this variable. Were this to be possible one could infer the photon spectral index at the source (See. Fig~\ref{figplusminus}).

%%%%%

\section{BSM Physics Sensitivity}
%%%%%
{Given the incredibly large energies and distances traveled, high-energy cosmic neutrinos can be used to probe new physics beyond the Standard Model (BSM).  We here provide three examples: neutrino decay, pseudo-Dirac neutrinos, and neutrino self-scattering.}  

\subsection{Neutrino Decay}
First, let us consider the effect of neutrino decay. Given the presumably large distances the neutrinos have travelled, high-energy neutrino telescopes may be uniquely sensitive to astrophysical neutrino decay~\cite{Beacom:2002vi}.  If the SM is only minimally extended to accommodate neutrino masses, then they are essentially stable on the timescales relevant for IceCube.  
{However in some BSM extensions such as Majoron models} neutrinos can decay either into the lightest neutrino or into other states, {and this possibility has been investigated before high-energy cosmic neutrinos are discovered} (e.g.,~\cite{Beacom:2002vi,Maltoni:2008jr,Baerwald:2012kc,Dorame:2013lka}).  
{Using the flavor information, Refs.~\cite{Pagliaroli:2015rca,nudecay} studied constraints on neutrino decay in light of the IceCube data on the diffuse neutrino flux.  In this work, using both flavor and spectral information, we demonstrate how next-generation detectors such as the {\it IceCube-Gen2} can be used to constrain neutrino decay.}  For simplicity, here we will assume that the decaying neutrinos have a large branching into some other non-interacting states. 
%\com{(This paragraph may include some more references and motivations.)}

The relevance of neutrino decay for a neutrino telescope is that each mass eigenstate has a given flavor fraction and can therefore modify the predictions for the number of cascades, tracks, Glashow resonance and double-bang events.  Moreover there is an important dependence on the neutrino mass.  This can be understood as follows. Given that the cosmic neutrinos are ultra-relativistic the decay rate of a given mass eigenstate depends on their relative boosts, $\Phi_{\nu_{i}} \propto \exp \left[-t (\gamma \tau_{i})^{-1}\right] =  \exp \left[-Lm_{i} (E_{\nu} \tau_{i})^{-1}\right]$, where $\tau_{i},m_{i}$ are the mass and rest-frame lifetime of mass eigenstate $i$. Therefore since the source-detector distance $L$ and neutrino energy $E$ pertain to the details of the source(s), the effects of neutrino decay are commonly quoted in terms of the parameter $\kappa_{i}^{-1} \equiv \tau_{i}/m_{i}$. {For simplicity, we will focus on a ``complete invisible decay'' model, in which the heaviest two mass eigenstates decay into invisible non-interacting states (e.g. sterile neutrinos).}

In the presence of decays the earthly flavor ratios can be estimated {\it schematically} as
\be
\alpha_{i}^{\oplus}(E) = \sum_{j,k}  \alpha_{j}^{S} |U_{ik}|^{2} |U_{j k}|^{2}~\exp \left[-\kappa_{k} L/E \right],
\label{eq:decayflavor}
\ee
where $L$ is the look-back distance and $E$ is the detected neutrino energy.  We follow the methods used in~\cite{Baerwald:2012kc} to estimate the effects from neutrino decay, which include the redshift-dependence of decay which are not included in Eq.~(\ref{eq:decayflavor}).
Here we assume that the sources follow the star-formation rate~\cite{Hopkins:2006bw,Yuksel:2006qb}.

The most stringent constraint on neutrino decay comes from the observation of SN1987A~\cite{Hirata:1987hu,Bionta:1987qt}. The 1987A constraint implies, $\tau_{i}/m_{i} \gtrsim 10^{5}~{\rm s}/{\rm eV}$~\cite{Hirata:1987hu,Bionta:1987qt}. {However, given the uncertainties in the total neutrino flux, this constraint may be applied to either $\nu_1$ or $\nu_2$, and it is possible to consider the following situations, e.g., $\nu_2$ and $\nu_3$ completely decay into $\nu_1$.}  Thus, most conservatively, this is simply interpreted as implying that only one mass eigenstate needs to have arrived at the detector.  Weaker constraints from solar~\cite{Joshipura:2002fb,Beacom:2002cb,Bandyopadhyay:2002qg}, atmospheric and long-baseline~\cite{GonzalezGarcia:2008ru} data apply to the other mass, but do not significantly constrain the parameter space that we are interested in for IceCube.

At the simplest level, we can consider the extreme case in which only a single mass eigenstate arrives at the detector at all energies. In this case, we can see from Fig.~\ref{fig111} that if equal flavor ratios are found at {\it IceCube-Gen2} then complete decay of either $\nu_{1}$ or $\nu_{3}$ would be strongly disfavored from the data ($> 4 \sigma$). See e.g. Figs. 1 and 5 of Ref.~\cite{Bustamante:2015waa} for the allowed flavor ratios of each mass eigenstate. Also note that a detailed look at present IceCube sensitivity to neutrino decay is forthcoming~\cite{nudecay}.

\begin{figure*}[t]
\begin{center}
  \includegraphics[width=.40\textwidth]{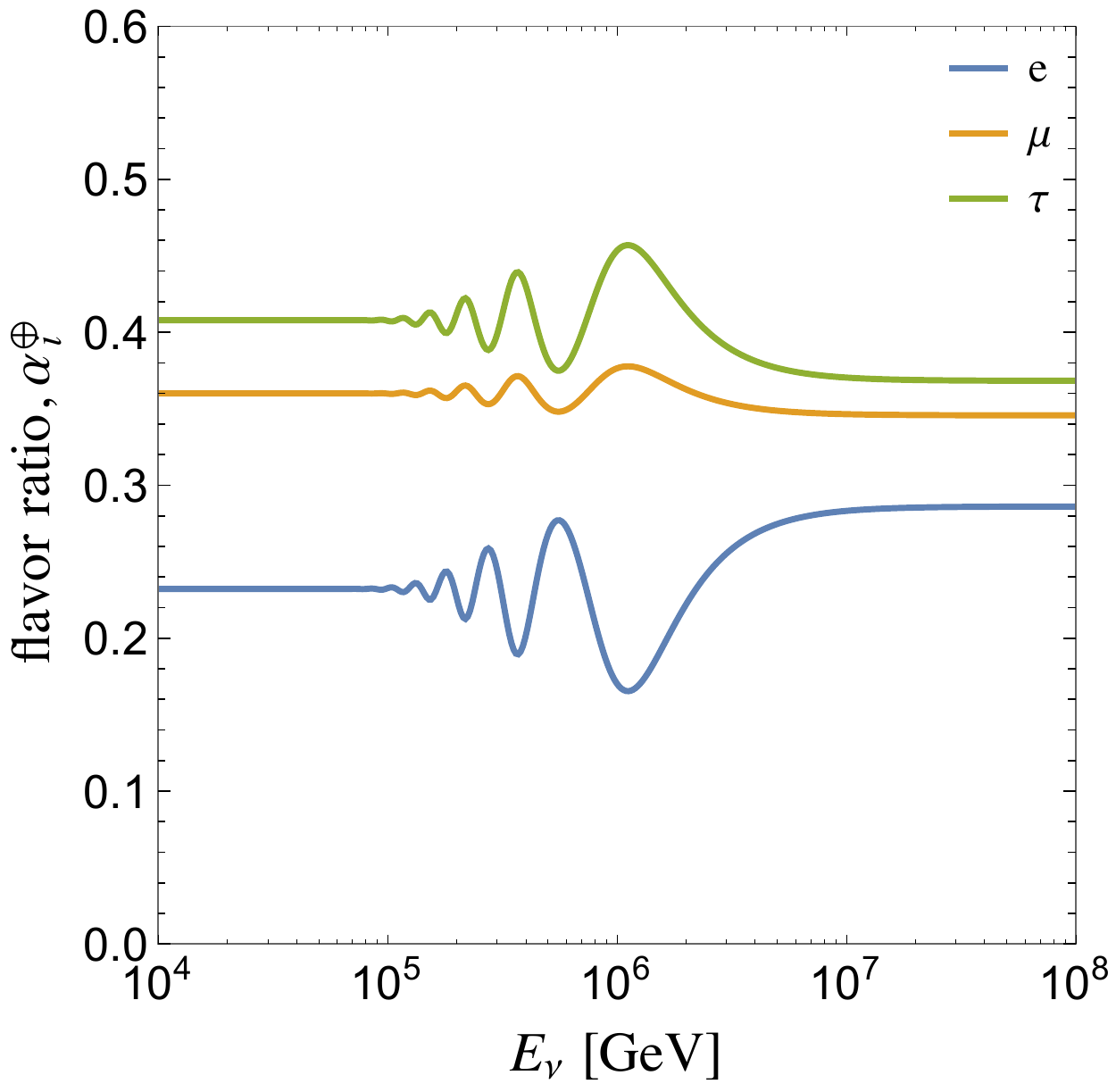} ~~~
  \includegraphics[width=.45\textwidth]{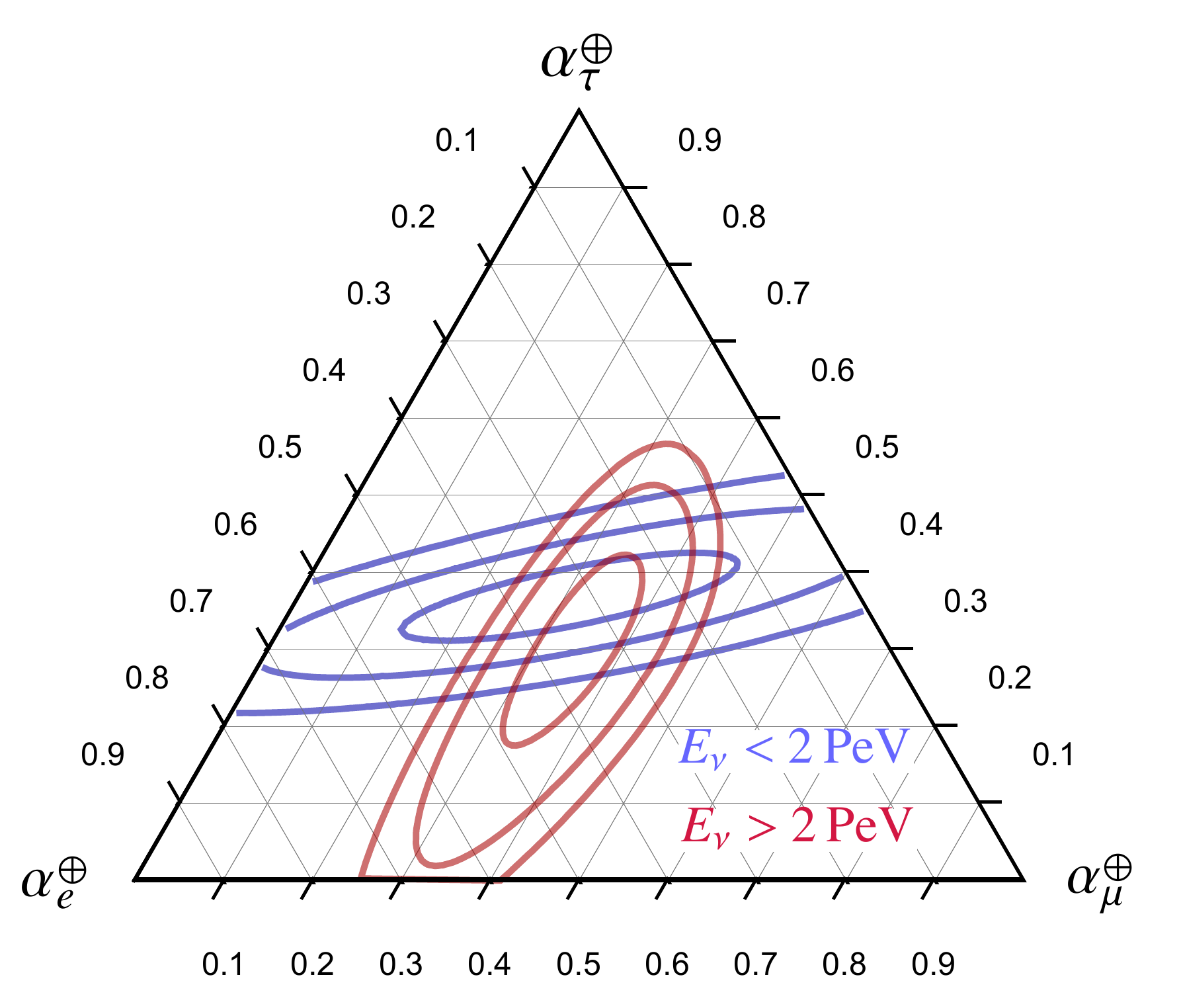} ~~~
\caption{Here we show an example of the flavor distortions that can arise from pseudo-Dirac neutrinos with a mass-splitting: $\Delta m_{k}^{2} = 10^{-17}~{\rm eV}^{2}$, with $k=1$ where we have assumed only one pseudo-Dirac neutrino split off the $\nu_{1}$ state. {The star-formation rate is used as a redshift evolution of the sources.}} %The right panel corresponds to mixing with 2 and 3 with $5\times 10^{-17}$ 
\label{pseudoDirac}
\end{center}
\end{figure*}

Next, consider the case of an ``incomplete decay'' in which only one mass eigenstate is present at the lowest energies but the flux transitions to the original source flavor ratios at higher energies. This most striking example of this is afforded in inverted hierarchy (IH) where only $\nu_{3}$ is stable. As displayed in Fig.~\ref{nudecay}, this depletes the $e$-flavor content at low-energies while leveling out to the standard (``undecayed'')  flavor ratios at high energies. Here we have taken $\tau_{1}/m_{1} = 10^{2}~{\rm s}/{\rm eV}$.

In order to empirically uncover the energy-dependent flavor induced by neutrino decay, we consider a flavor fit in two different energy bins: above $2$ PeV and below 2 PeV. The result of these two fits is depicted in Fig.~\ref{nudecay} where we demonstrate that an energy dependent flavor determination is possible. %In the high energy bin ($E_{{\rm dep}} > 2~{\rm  PeV})$ we get 54 showers, 4 tracks, 17 Glashow resonance events, and 5 double-bangs. In the low-energy bin (0.1 PeV $< E_{{\rm dep}}  < 2$ PeV) we get 109 showers and 37 tracks.  
We note that this example may be in a mild $(1-2)\sigma$ tension with the current flavor constraints from combined maximum likelihood analysis of IceCube's events~\cite{Aartsen:2015knd}. 
{In Sec.~\ref{sec:discuss}, we will show that future neutrino detectors such as {\it IceCube-Gen2} can provide us with more stringent constraints on neutrino decay through a joint flavor and spectral analysis.}  

%%%%
\subsection{Oscillating into New States: Pseudo-Dirac Neutrinos}
%%%%
The nature of origin of neutrino masses remains poorly understood, but many models predict the existence of right-handed sterile neutrinos. These states have of course been searched for in a number of realms. The well-known seesaw mechanism predicts that these states have very large Majorana masses that make them otherwise hard to probe. By contrast in the pseudo-Dirac scenario, the Majorana masses are small compared to the Dirac scale, and the induced small mass-splittings provides another mechanism which makes right-handed neutrinos hidden from us.
{We here consider the effect of the pseudo-Dirac neutrinos~\cite{Wolfenstein:1981kw,Petcov:1982ya}, in which there may exist a tiny mass splitting between the active and sterile neutrinos.  
Applications to astrophysical neutrinos have been considered in Refs.~\cite{Beacom:2003eu,Keranen:2003xd} (see also Ref.~\cite{Esmaili:2009fk}) before high-energy cosmic neutrinos discovered.} 

These small mass splittings only gives rise to oscillations to the sterile state on very large distance scales, since the oscillation length is
\be
L_{{\rm osc}} = 80~{\rm Mpc}~\left(\frac{E}{1~{\rm PeV}}\right)~\left(\frac{10^{-15}~{\rm eV}^{2}}{\Delta m_{j}^{2}}\right),
\ee
where $\Delta m_{j}^{2}$ is the mass-splitting with the $j$th active neutrino mass eigenstate.

In this case, the neutrino flavor ratios at the Earth can be very different and depend sensitively on the energy: 
\be \alpha_{i}^{\oplus} = \sum_{j,k}  \alpha_{j}^{S} |U_{ik}|^{2} |U_{j k}|^{2}~\cos^{2}\left(\frac{\Delta m^{2}_{k} L}{4E}\right).
\label{eq:pd}
\ee
In the above $L$ is the distance between the source and the Earth. Notice that one recovers the pure Dirac result in the vanishing $\Delta m^{2}_{k}$ limit. 
These new mass splittings could be off of only one of the active neutrinos or off all of them. The mass splittings with $\nu_{1}$ has the largest effect though, as it induces a large effect on the electron-flavor component. 

Notice that Eq.~(\ref{eq:pd}) has two shortcomings: (1) it assumes implicitly a static Universe, and (2) it assumes a single source at a given distance from the observer.  
{The first point can be easily addressed by computing the phase difference in an expanding Universe.  The proper phase difference is calculated as~\cite{Esmaili:2012ac}
\be \Delta \Phi_{j} = \frac{\Delta m_{j}^{2}}{2E} D_{H} \int_{0}^{z} \frac{dz'}{(1+z')^{2} \sqrt{\Omega_{m}(1+z')^{3}+ \Omega_{\Lambda}}}.
\ee
with $D_{H} =H_{0}/c$ the Hubble distance and $\Omega_{m} = 0.27$  and $\Omega_{\Lambda} = 0.73$. Then, according to the second point, one needs to consider a population of sources tracing a known rate distribution such as the star-formation rate. Then, the neutrino flavor ratios at the Earth becomes
\be
\alpha_{i}^{\oplus} = \sum_{j,k}  \alpha_{j}^{S} |U_{ik}|^{2} |U_{j k}|^{2}~\left \langle \cos^{2}\left(\frac{\Delta \Phi_{j}}{2}\right) \right \rangle. 
\label{eq:pd2}
\ee
The angled brackets in Eq.~(\ref{eq:pd2}) denote energy average over the resolution of the detector which is assumed to follow a Gaussian energy distribution with resolution $\sigma_{E} = 0.15 E$.}
Then, we include the effect of source distribution as in~~\cite{Esmaili:2012ac} and assume that they track the star-formation rate~\cite{Hopkins:2006bw,Yuksel:2006qb}.

We display this behavior in Fig.~\ref{pseudoDirac}. We see that as in the neutrino decay case, there is sufficiently good sensitivity to reconstruct some aspects of the energy-dependence of the flavor ratios, though in this case not quite as efficiently as in the case of neutrino decay. Moreover though flavor properties such as these would indicate the presence of some new BSM physics in the neutrino sector, distinguishing neutrino decay from pseudo-Dirac neutrinos will be challenging. A more optimistic path for discrimination {\it between} BSM scenarios will be offered by a joint flavor and spectral analysis ({see Sec.~\ref{sec:discuss}}).  

\begin{figure*}[t]
\begin{center}
  \includegraphics[width=.45\textwidth]{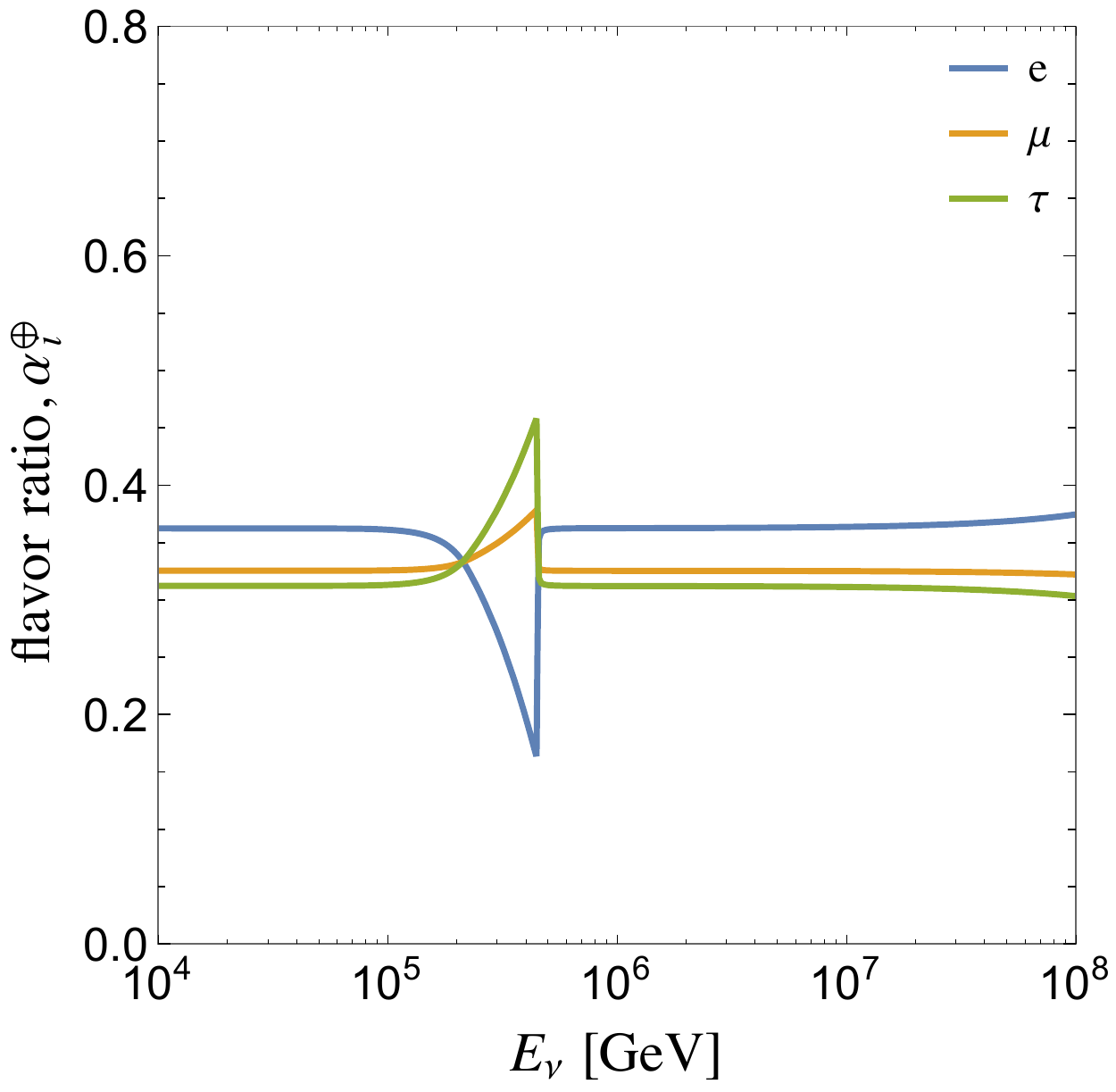} ~~~
  \includegraphics[width=.45\textwidth]{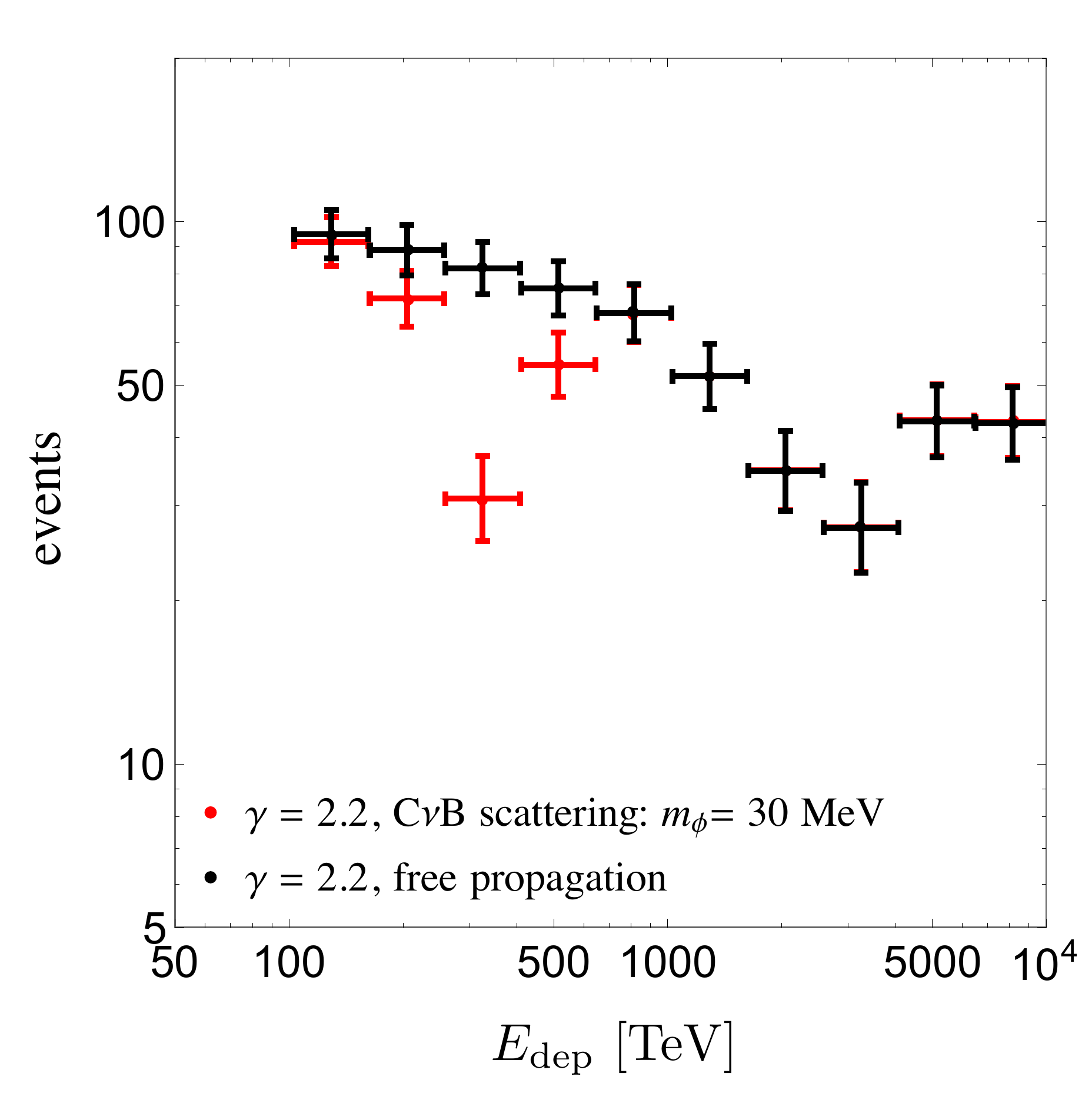} ~~~
\caption{Here we show an example of the spectral distortions that secret interactions can engender. The black data points show a free-propagation (i.e. no self-scattering) example with a $\gamma =2.2$ spectrum and equal flavor ratios. The red data points show the effect that a $m_{\phi} = 30$ MeV mediator can have from scattering on the C$\nu$B. Here we have taken the coupling $g = 0.1$ and mixing angle $\sin \theta =0.1$.}
\label{secret}
\end{center}
\end{figure*}

%%%%%%%%%
\subsection{Neutrino Self-Scattering}
%%%%%%%%%
Lastly, we consider the effect of neutrino self-scattering~\cite{BialynickaBirula:1964zz,Bardin:1970wq} on the Cosmic Neutrino Background (C$\nu$B) en route between the astrophysical source and the Earth.  We assume that astrophysical neutrino source produces only some combination of the active flavor ratios, though the scattering partners in the C$\nu$B can be either active or sterile neutrinos. The large number density of relic neutrinos in the C$\nu$B, $\sim 100~{\rm cm}^{-3}$, makes sizable neutrino self-scattering a possibility if they interact with new forces, sometimes called ``secret interactions'' and applications to cosmic neutrinos have been considered~\cite{Kolb:1987qy,Manohar:1987ec,Keranen:1997gz}.
{Soon after cosmic high-energy neutrinos were discovered by the IceCube Collaboration, it was pointed out that the IceCube data can be used as an unique probe of the secret interactions of neutrinos~\cite{Ioka:2014kca,Ng:2014pca}, and some detailed models have been constructed~\cite{Ibe:2014pja,Blum:2014ewa,Araki:2014ona,Cherry:2014xra,Kamada:2015era,DiFranzo:2015qea}.}  

One of the simplest ways to achieve the requisite cross sections for significant scattering is through the resonant exchange of mediator particle.  We will refer to this mediator simply as $\phi$ though it could be a scalar~\cite{Ioka:2014kca,Ng:2014pca,Blum:2014ewa} or a vector~\cite{Aarssen:2012fx,Harnik:2012ni,Laha:2013xua,Cherry:2014xra} boson.  Note that in models with direct couplings to active neutrinos, a number of laboratory constraints exist~\cite{Harnik:2012ni,Laha:2013xua,Blum:2014ewa}. These bounds are considerably relaxed if the mediator only couples to sterile neutrinos, since flavor transitions need to occur inside the detector with large probability. To our knowledge there is no detailed study of how the constraints change in a model with couplings only to the sterile state.

%\com{(Note that there are constraints from experiments. The vector meson coupled to active neutrinos is constrained to be $g\sim0.03$ but it should be weaker for the meson only coupled to sterile neutrinos, as we discussed. However, to see the effect of neutrino attenuation we need to relatively large values of $g^2\sin\theta$ (including the mixing angle). So, eventually, are these constraints competing if you want to have observable effects in IceCUbe? Possibly it may be good to show or discuss constraints from Z-decay ($Z\rightarrow \nu\nu\phi$) and kaon-decay ($K\rightarrow \nu\mu\phi$).)}
When the relic C$\nu$B scattering partner is very non-relativistic ($m_{\nu} \gg T_{{\rm C \nu B}} = (4/11)^{1/3}T_{\rm CMB} \simeq 1.7 ~{\rm meV}$), we expect a significant depletion of the flux at the resonant energy
\be E_{\rm res} = \frac{m_{\phi}^{2}}{2m_{\nu}} \simeq 0.5~{\rm PeV}~\left(\frac{m_{\phi}}{10~{\rm MeV}}\right)^{2}~\left(\frac{0.1~{\rm eV}}{m_{\nu}}\right)
\ee
where $m_{\phi}$ is the mass of the mediator $\phi$. We see that mediator masses at the MeV-scale can impact PeV neutrino propagation. Mediators in this mass range have also been suggested to help reconcile eV sterile neutrinos with cosmology~\cite{Dasgupta:2013zpn,Hannestad:2013ana} and to alleviate problems with the small-scale structure anomalies of collisionless dark matter~\cite{Loeb:2010gj,Aarssen:2012fx,Dasgupta:2013zpn,Cherry:2014xra}.
Note that while $t$-channel processes to the scattering contribute as well they are only relevant in the case of relatively large coupling which is not allowed in the case of active neutrino coupling~\cite{Ioka:2014kca,Blum:2014ewa,Kamada:2015era,DiFranzo:2015qea} but can be important for sterile neutrino couplings despite the mixing angle suppression~\cite{Cherry:2014xra}.
 
In the following we will restrict ourselves to absorption of active neutrino flux without regenerating it at lower energies. Two physical realizations of this are when the produced mediator has a dominant branching ratio into sterile neutrinos (e.g.,~\cite{Cherry:2014xra}) (Model 1) or Nambu-Goldstone bosons (e.g.,~\cite{Blum:2014ewa}) (Model 2).  
In the alternative case, where the mediator decays back to active neutrinos the overall effect is to remove flux at high energy and deposit it back at lower energy, leading to a ``pile up'' of flux compared to the un-scattered flux.  {For the effect of neutrino cascades via secret interactions among active neutrinos, see Refs.~\cite{Ioka:2014kca,Ng:2014pca,Blum:2014ewa}.}

\begin{figure*}[t]
\begin{center}
  \includegraphics[width=.32\textwidth]{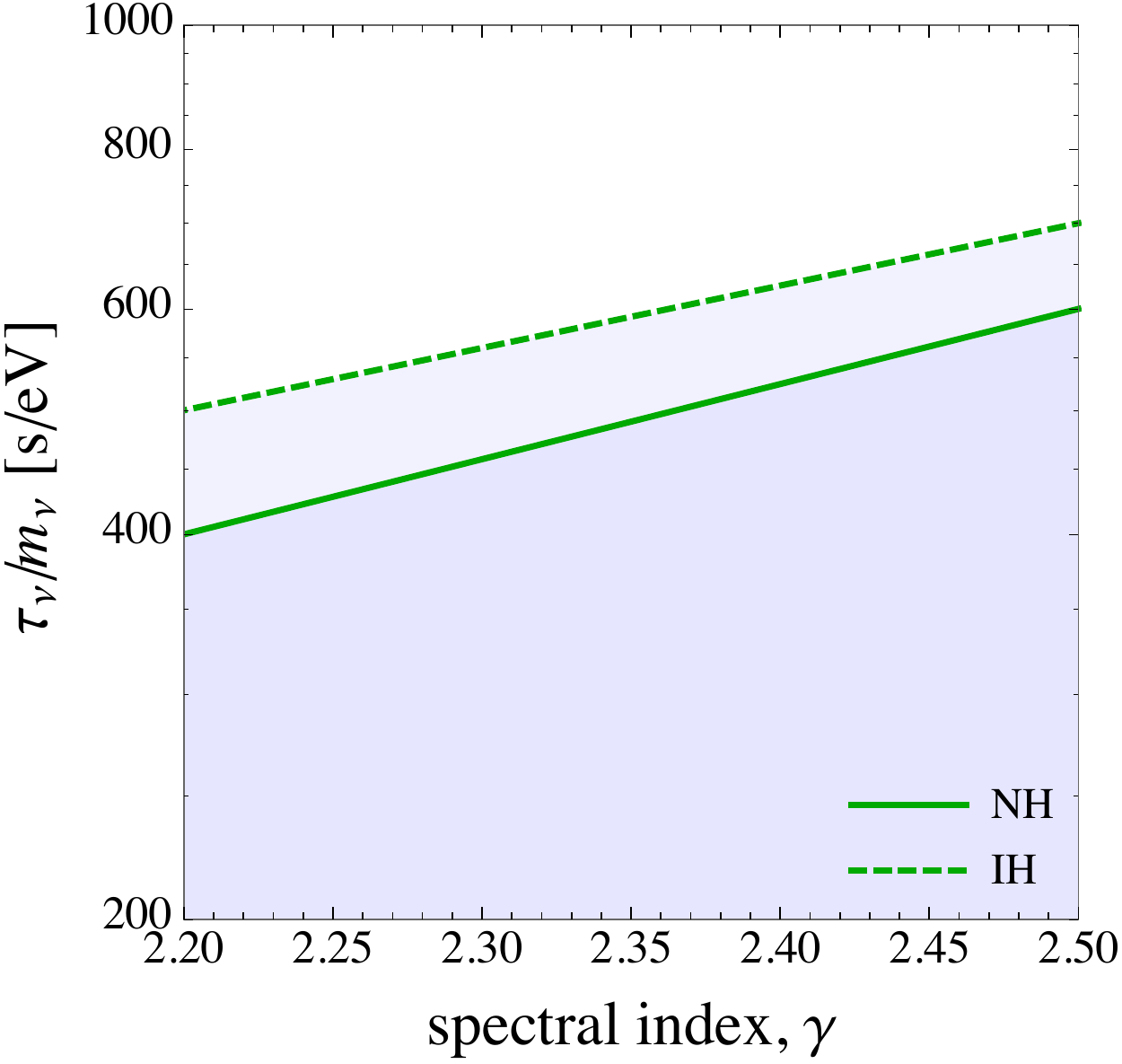} ~
  \includegraphics[width=.32\textwidth]{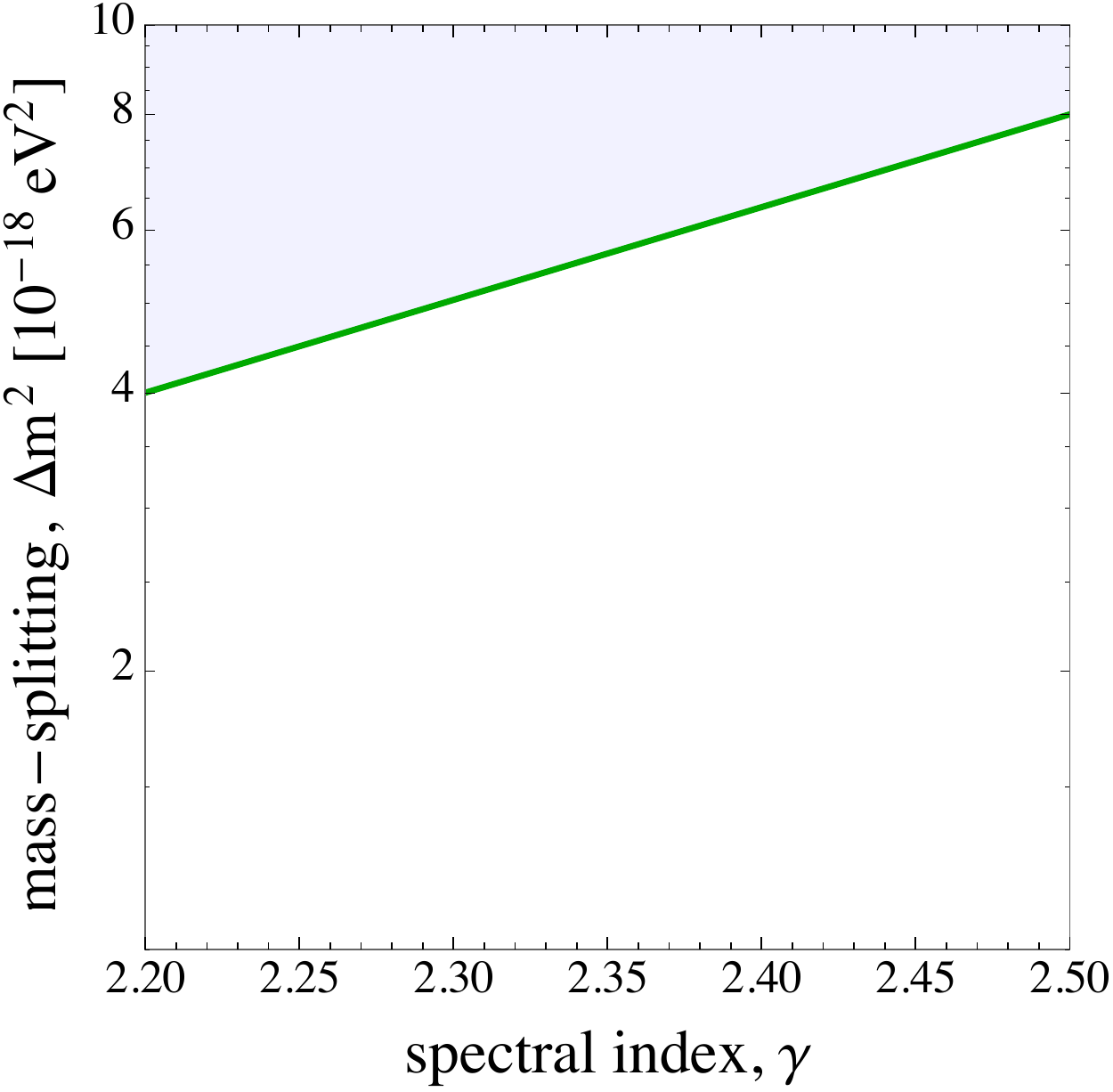} ~
    \includegraphics[width=.32\textwidth]{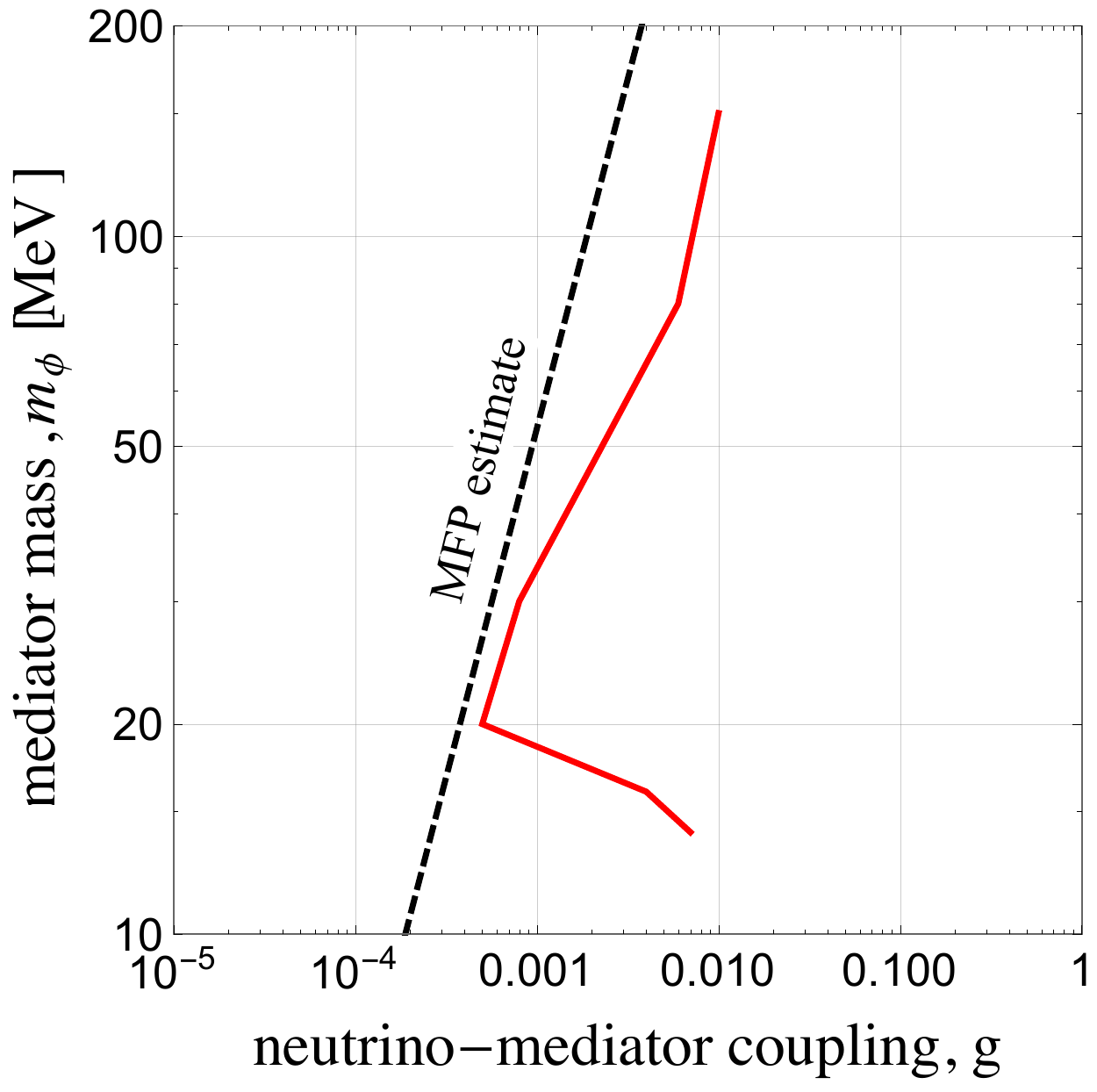} ~
\caption{
Here we summarize the expected the $2\sigma$ sensitivity to BSM neutrino at a future experiment like the proposed {\it IceCube-Gen2 } detector with 10 years of data. 
{\bf Left Panel:} The forecasted constraints on neutrino decay as a function of the source spectral index. We have assumed $pp$ source flavor ratios but have verified that the results are not very different in $p\gamma$ scenarios. 
{\bf Middle Panel:} The forecasted constraints on Pseudo-Dirac neutrinos. 
{\bf Right Panel:} The forecasted constraints on neutrino secret interactions ({see the text}). 
Here we have fixed the spectral index to, $\gamma =2.2$. Note that the lower boundary is not {shown} because the spectral distortions {lie below the 100 TeV energy threshold, where the atmospheric backgrounds are relevant. On the other hand, cases of very heavy mediator mass are not shown from the fact the the spectral distortions only modify high-energy bins which have no events.}
}
\label{summary}
\end{center}
\end{figure*}

To include the effect of {neutrino attenuation due to their} self-scattering at different redshifts and the source redshift dependence we solve the differential equation for the neutrino spectrum ($f_{i} \equiv dn^{\rm HE}_{i}/dE_{\nu}$) of mass eigenstate $i$ at the Earth 
\be
-(1+z) H(z) \frac{d f_{i}}{dz} = J_{i}(E',z) - f_{i} \sum_{j} \langle n_{\nu_{j}} \sigma_{ij}(E',z) \rangle,
\ee
where $n_{\nu_{j}}(z) = n_{\nu,0} \left(1 + z\right)^{3}$, $\sigma_{ij}(E',z)$ is the cross section for mass eigenstates $ij$ scattering to non-interacting states, and $H(z) = H_{0} \sqrt{\Omega_{\Lambda} + (1+z)^{3} \Omega_{m}}$.  Lastly, {$J_{i}(E',z) = \phi_{i} {E'}^{-\gamma} f(z)$ is the source distribution function where we take the redshift distribution $f(z)$ to follow the star formation rate~\cite{Hopkins:2006bw,Yuksel:2006qb}}.

Now we illustrate in Fig.~\ref{secret} the effect that absorption features from secret interactions can have at {\it IceCubeGen-2}. First, notice that the flavor modifications are restricted to a narrow range of energies, making it challenging with only flavor information to determine much since the statistical power over a small energy range is limited.  The fact that only a small range of energy is effect can be understood as a consequence of assuming resonant scattering on a non-relativistic target. The neutrinos could instead be relatively warm, either from having a larger temperature than naively expected or by having mass smaller than or comparable to the $T_{{\rm C\nu B}}$. Then in this case thermal broadening would allow for a significant scattering over a much wider range of energies~\cite{Barenboim:2004di,Lunardini:2013iwa}. Note that for the case displayed in Fig.~\ref{secret} the resonance feature is highly asymmetric (see also Ref.~\cite{Blum:2014ewa}). This is a direct consequence of redshifting of the C$\nu$B, with the high-$z$ scattering leading to the largest suppression (since the neutrino density scales as $\propto (1+z)^{3}$)~\cite{Barenboim:2004di}.  

If future data uncovers a spectral distortion like the red data points in the left panel of Fig.~\ref{secret} {\it IceCubeGen-2} will be able to exclude the hypothesis of a pure power law without neutrino scattering at $>3 \sigma$. Notice that although we do not attempt to perform a full scan of the parameter space we anticipate that similar statement can be made for secret interactions that give rise to a resonant absorption feature anywhere below the Glashow resonance (and above the atmospheric background). 

Although we have stressed that absorption models like Ref.~\cite{Cherry:2014xra} (Model 1) and Ref.~\cite{Blum:2014ewa} (Model 2) lead to qualitatively similar effects, we have chosen to display in Fig.~\ref{secret} an example of Model 1. Here we have assumed that the incoming high-energy neutrinos scatter on a eV sterile neutrino~\cite{Cherry:2014xra} that mixes with $\nu_{1}$ with a vacuum mixing angle $\sin \theta =0.1$. At low temperatures the combination of oscillations and collisions leads to a ``recoupling'' between the sterile and active neutrinos, thereby giving them comparable densities and temperatures~\cite{Bringmann:2013vra,Mirizzi:2014ama,Cherry:2014xra,Chu:2015ipa}.
Upon scattering with the $\nu_{4}$ states in the C$\nu$B the mediator decays back to sterile neutrinos which are dominantly composed of $\nu_{4}$. We note that the results of Fig.~\ref{secret} could equally well apply to the model of active neutrino self-scattering model in~\cite{Blum:2014ewa}, with the replacements
\be 
m_{\phi,{\rm Model~2}} = m_{\phi,{\rm Model~1}}~\sqrt{\frac{m_{\nu_{\alpha}}}{m_{\nu_{s}}}},
\ee
which accounts for the shift in the resonance energy. Furthermore, the additional factor of $\sin^{2} \theta$ in the sterile neutrino case can roughly be accounted for by rescaling couplings: $g_{{\rm Model~2}} \simeq g_{{\rm Model~1}} \sqrt{ \sin \theta}$. Note that we define the coupling in each model as ${\mathscr L}\supset g \nu\nu\phi$. 
%(\com{Could you check if the coefficient is correct? Clarify how the coupling is introduced (I think ${\mathscr L}\supset g \nu\nu\phi$). })

\begin{figure*}[t]
\begin{center}
  \includegraphics[width=.45\textwidth]{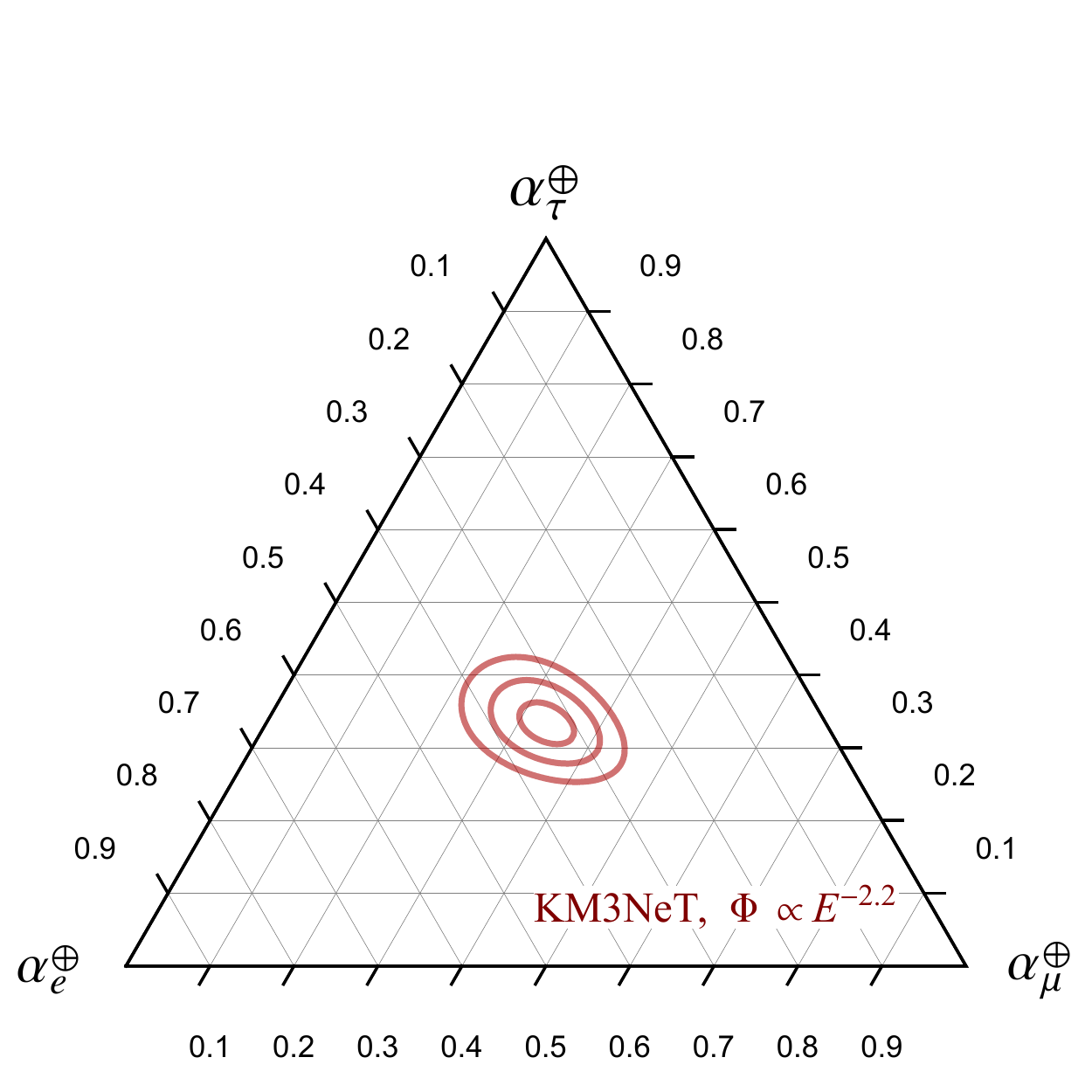} ~~~
  \includegraphics[width=.45\textwidth]{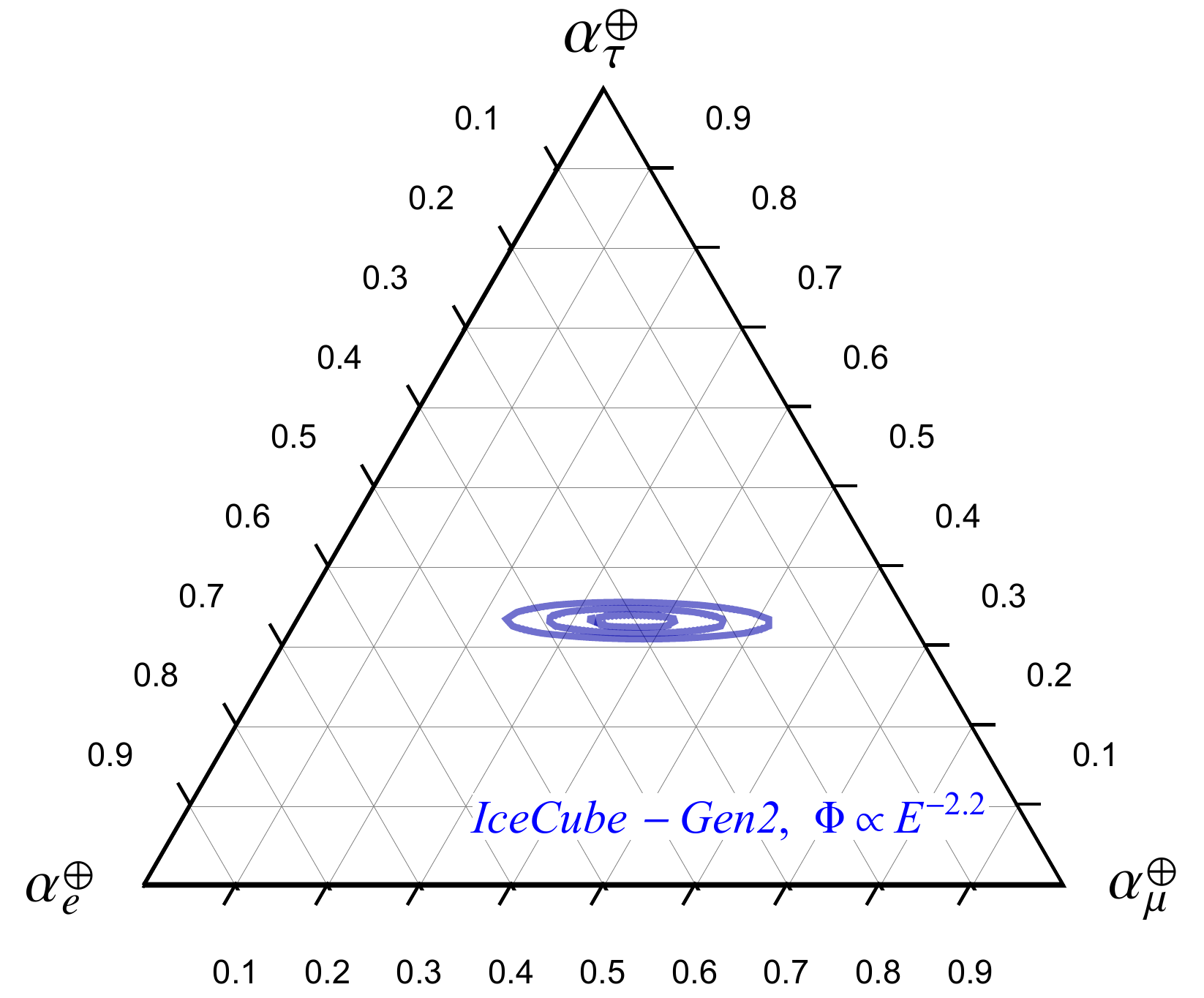} ~~~
\caption{As can be seen, with an optimistic ``theorist's approach' analysis, next-generation neutrino detectors such as KM3NeT~\cite{Adrian-Martinez:2015rtr} and {\it IceCube-Gen2 } will have significantly better flavor sensitivity with {dedicated analyses combining shower and track events.} As in all plots, we have assumed a 10 year exposure.} %The right panel corresponds to mixing with 2 and 3 with $5\times 10^{-17}$ 
\label{nuflavor}
\end{center}
\end{figure*}

Lastly, notice a degeneracy which arises when the absorption feature occurs near the Glashow resonance at $\sim 6.3$ PeV. This can eliminate the main source of electron flavor discrimination. On the one hand, we can see that from Table~\ref{table} a $pp$ source plus Glashow-suppressing secret interactions may be difficult to distinguish from a $p\gamma$ without secret interactions. It would also be challenging to distinguish from a pure $\mu$ and $\tau$ terrestrial flux which could arise from a neutrino decay in which only $\nu_{3}$ is present at Earth. A situation such as this could significantly hamper source discrimination. 

%We stress that this is a conservative example of neutrino self-scattering and note that if the neutrinos are still relativistic much more extreme possibilities with absorption over a wide energy range could be observed.
%KM: it is not clear if this is conservative. 

%CPT notes: 1001.4878
%%
\subsection{Discussion}
\label{sec:discuss}
%%%%
We here summarize the sensitivity to the BSM modifications we have considered in Fig.~\ref{summary}. In all cases we estimate the sensitivity that an experiment like the {\it IceCube-Gen2 } detector will have with a HESE-like analysis by varying the spectral index, $\gamma$. 

The left figure examines neutrino decay assuming only that the $\nu_{1}$ is stable (NH) and only that $\nu_{3}$ is stable (IH). Slightly stronger limits are likely in the IH simply because $\nu_{3}$ has almost no electron flavor content which makes this flavor signature quite distinct. The center panel examines the sensitivity to pseudo-Dirac neutrinos where the new state is split off of $\nu_{1}$ (e.g. Fig.~\ref{pseudoDirac}). We note that here spectral feature play a more dominant role. We verified this explicitly by examining the case that each active mass eigenstate has a new pseudo-Dirac partner, in which case the flavor ratios maintain their $pp$ values at all energies even though there is an all-flavor spectral distortion. In the final case of secret interactions we examine Model 1. Here we assume the modifications arise from scattering on the sterile neutrino component of the C$\nu$B and have fixed the spectral index to $\gamma =2.2$. At sufficiently high mediator masses only the very high energy portion of the spectrum is affected. At a critical mediator mass this means that with a given exposure, neutrino self-scattering only modifies energy bins which do not yet contain any events. The defines the upper boundary of the figure. The lower figure is set by the 100 TeV energy threshold we have assumed. In principle constraints exist in this region as well from scattering on the lighter active mass eigenstates.

Of course, there are a large number of BSM scenarios that {\it IceCube-Gen2} will have sensitivity to that we have not yet discussed. For example, CPT violation in the neutrino sector remains possible and can effect the flavor and spectra of high-energy neutrinos~~\cite{Barenboim:2003jm,Kostelecky:2003xn,Hooper:2005jp,Anchordoqui:2005gj,Dighe:2008bu,Ando:2009ts,Bustamante:2010nq,Arguelles:2015dca} (see also~\cite{Diaz:2013wia,Stecker:2014oxa,Anchordoqui:2014hua} for implications of Lorentz violation).  One signature of this could come from neutrinos coupling to the CPT violating piece, $\mathscr{L}_{CPTV} \supset b^{\mu}_{\alpha \beta} \bar{\nu}^{\alpha} \gamma_{\mu} \nu_{\beta}$, where $ b^{\mu}_{\alpha \beta} $ is a constant flavor-dependent vector. The combination of solar and atmospheric data constrain these to satisfy $b_{21} < 1.6 \times 10^{-21}~{\rm GeV}$ and $b_{32} < 5.0 \times 10^{-23}~{\rm GeV}$.  Roughly speaking we should expect sizable modifications from CPT violation of this type if the coefficients are of order the ordinary oscillation scale $\simeq \Delta m^{2}/2 E$. We therefore anticipate that {\it IceCube-Gen2} will be able to probe such CPT violating effects down to the level, $b_{21} \simeq 10^{-29}~{\rm GeV}^{-1}$ and $b_{32} \simeq 10^{-28}~{\rm GeV}^{-1}$. 

Conceivably with these very large exposures, {\it IceCube-Gen2} can also be sensitive to the directionally dependent oscillations that can be induced by a neutrino-dark energy coupling~\cite{Ando:2009ts}.  Such an effect depends on the neutrino's direction relative to the Earth's peculiar velocity with respect to the CMB. Another interesting modification to standard physics is the presence of extra dimensions~\cite{Pas:2005rb}. This could impact neutrino oscillations since the amplitude of flavor conversion since the path length for sterile oscillations is modified. Models of this type can result in strongly energy-dependent flavor distortions or more dramatic complete conversions when the extra-dimensional shortcut varies adiabatically~\cite{Aeikens:2014yga}. This latter case can result in a total depletion of {$\nu_\mu$ and $\nu_\tau$ flavors}. 

In our example of neutrino self-interactions we have focused on the case of coupling to an eV sterile neutrino.  Such a sterile neutrino may already be hinted at in short-baseline data and ~\cite{Aguilar:2001ty,Aguilar-Arevalo:2013pmq,Mueller:2011nm,Mention:2011rk,Hayes:2013wra,Acero:2007su} and will be searched for via the Short-baseline Neutrino Program at Fermilab~\cite{Antonello:2015lea}. Additional tests of the interacting sterile neutrino model may come from cosmology~\cite{Cyr-Racine:2013jua,Chu:2015ipa} and cross-correlation searches for the sources of the high-energy cosmic neutrinos~\cite{Cherry:2014xra}.

We have considered BSM physics that can be probed by utilizing astrophysical neutrinos {(see also e.g.~\cite{Bhattacharya:2009tx,Bhattacharya:2010xj,Anchordoqui:2013dnh} for a variety of BSM possibilities at neutrino telescopes)}. However, neutrinos may originate from heavy dark matter which would also require {BSM physics~\cite{Feldstein:2013kka,Rott:2014kfa,Esmaili:2013gha,Esmaili:2014rma,Ema:2014ufa,Fong:2014bsa,Anchordoqui:2015lqa,Murase:2015gea}}. In this case, the source flavor ratios can take any value, though the effect of neutrino oscillations narrows the allowed terrestrial flavor ratios considerably~\cite{Bustamante:2015waa}.

Lastly, we should emphasize that our analysis throughout the paper has been based on a HESE-like search strategy. Since {relatively tight cuts are applied in} these searches our results up to this point can be viewed as rather conservative. An obvious question is: What will a more optimistic analysis reveal? Since the efficiency of upgoing muon and {dedicated} shower analyses can be considerably larger than for HESE events, we expect a sizable gain in statistics by considering these event {cuts}.  We optimistically mock-up the sensitivity to these events by following the ``theorist's'' approach borrowed from~\cite{Laha:2013eev,Blum:2014ewa} which roughly speaking trades the effective area for the product of the neutrino cross section~\cite{Connolly:2011vc} and Earth-induced neutrino attenuation. 
Here we scale up the volume by a factor of 10 for a {\it IceCube-Gen2}-like detector. 
Moreover, since we expect KM3NeT~\cite{Adrian-Martinez:2015rtr} to be operational in the same timeframe, {we also include a simple mock-up of their sensitivity for which we assume a volume of $\simeq 1~{\rm km}^{3}$ assuming three blocks of KM3NeT. We also assume that the optical properties of water allow us to reduce a double-bang threshold to 10 TeV.}

We show our flavor fits under the above assumptions for {\it IceCube-Gen2} and KM3NeT in the left and right panels of Fig.~\ref{nuflavor} respectively.  As can be seen, these results look dramatically more constraining than in the HESE-like results of Fig.~\ref{fig111}. The gain is especially significant in the muon flavor direction, which follows as a consequence of the larger effective volume since {neutrinos may interact with ice} considerably outside of the detector contribute here. Note that since Fig.~\ref{nuflavor} should be taken as an optimistic projection, we have adopted a spectral index $\gamma=2.2$. 

%%%%
\section{Conclusions}
%%%%
We have seen that {next-generation neutrino telescopes such as the proposed {\it IceCube-Gen2} detector in Antarctica and KM3NeT detector in the Mediterranean Sea} will provide important flavor information on high-energy neutrino signal. First and foremost, this information can be used to reveal the origins of these neutrinos.  We have shown that the sensitivity of {\it IceCube-Gen2} is sufficient to detect both deviations in the individual flavor ratios and also in the $\nu$ and $\bar{\nu}$ flux ratios. This latter possibility is afforded entirely by the existence of the Glashow resonance effect which is sensitive to the $\bar{\nu}_{e}$ flux. Given the sizable statistics {\it IceCube-Gen2} will offer, we additionally estimated the ability to test new BSM physics scenarios.  In particular, neutrino decay, pseudo-Dirac neutrinos, and neutrino secret interactions can produce distinctive flavor and spectral signatures, and {we provided forecasted constraints in the {\it IceCube-Gen2} era.} {We saw that hadronuclear ($pp$) and photohadronic ($p\gamma$) source models can produce distinctive flavor ratios, although they can in principle be distinguished if the $\pi^{-}$ production is not significant in the photohadronic case.} We are in an exciting era where high-energy neutrino telescopes are gaining access to physical regimes that can test many BSM physics scenarios that are otherwise impossible to test.

\section*{Acknowledgements}
{We are very grateful to Markus Ahlers, John Beacom, Mauricio Bustamante, and Uli Katz for helpful comments and careful readings of the manuscript. We also thank Lars Mohrmann for provide us with the contour data of the combined likelihood analysis of Ref.~\cite{Aartsen:2015knd}.}  

%%%%
\section*{Appendix: Case of $p\gamma$ scenarios}
%%%%

\begin{figure}[b]
\begin{center}
 \includegraphics[width=.4\textwidth]{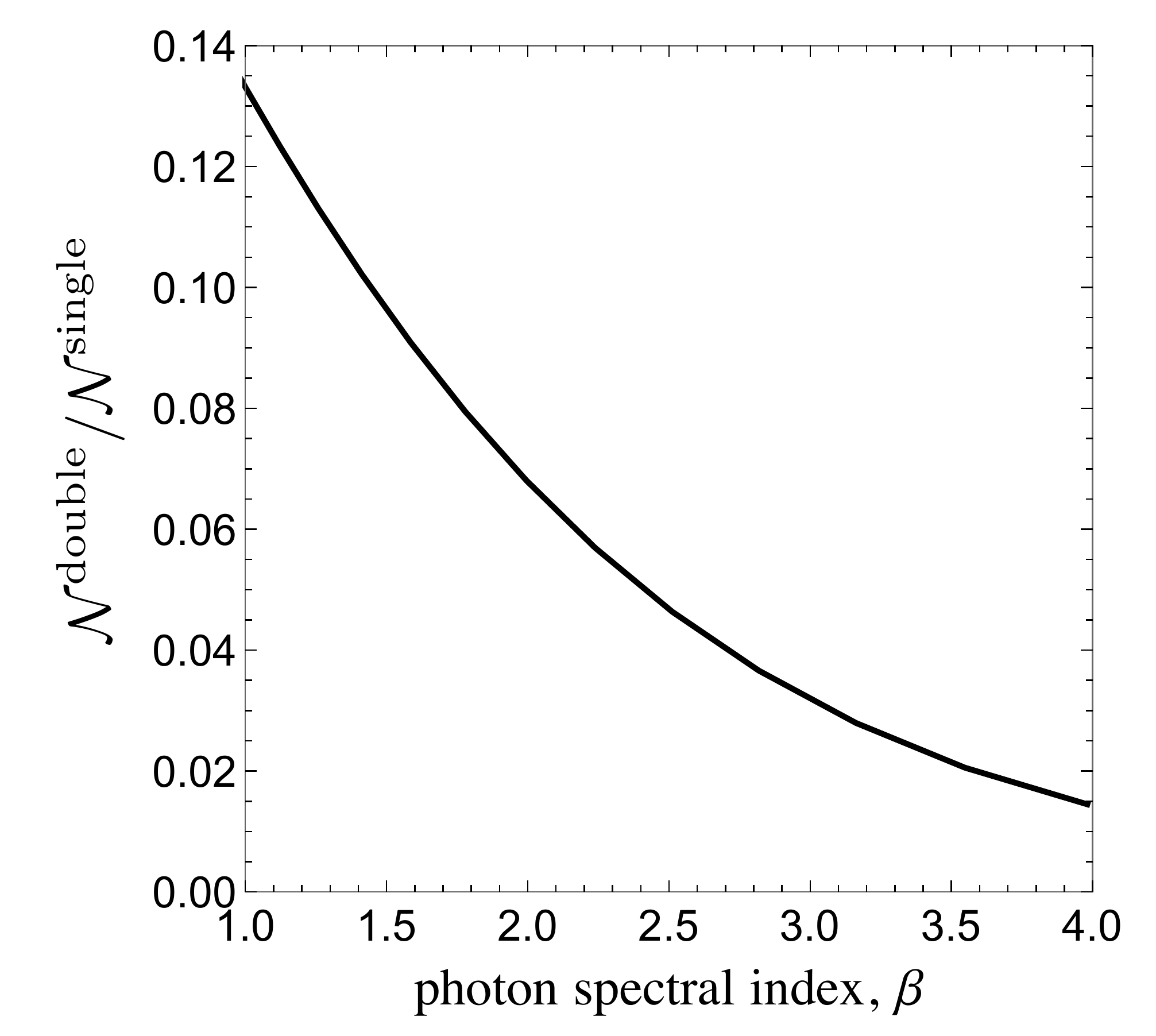} \\
   \includegraphics[width=.40\textwidth]{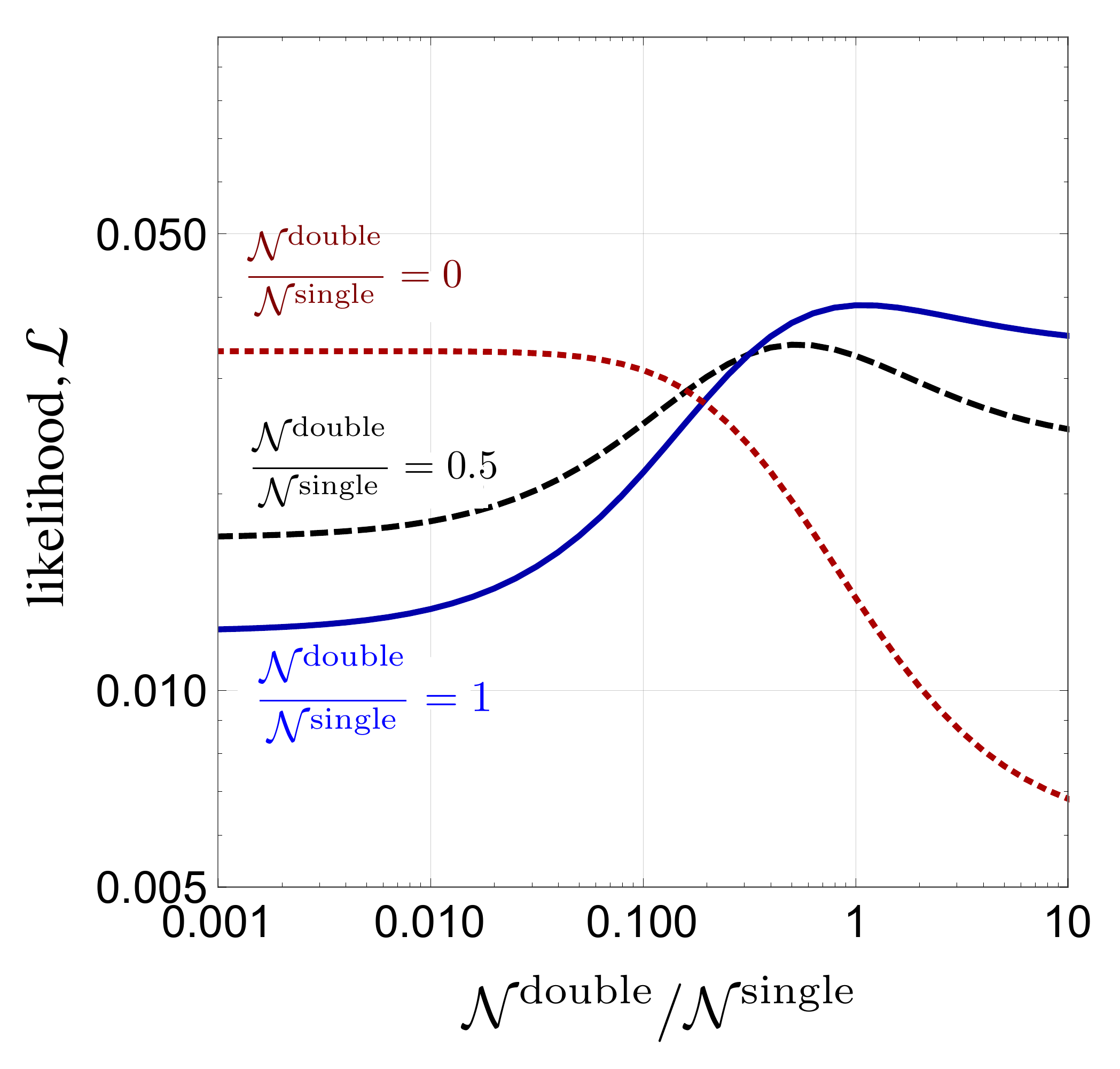}
\caption{ {\it Top panel:} Expected ratio of double-to-single pion production from photohadron source as a function of the target photon spectral index.
 {\it Bottom panel:} Here we demonstrate the role of both the threshold energy $E_{{\rm dep}}^{{\rm thr}}$ and the input spectral index $\gamma$ where the fits here include all flavor discriminants. Note that a detailed background study may alter the 10 TeV threshold results.}
\label{figplusminus}
\end{center}
\end{figure}

For completeness we include in this Appendix some additional details regarding our computation of the flavor ratios in $p\gamma$ scenarios. 
Weighting each process's contribution to the total source flavor ratio by their average cross sections~\cite{Schadmand:2005ji} give the energy-averaged flavor ratios
\be
\alpha_{S}\approx\alpha_{S}^{(0)} + \alpha_{S}^{(1)}   \frac{{\mathcal N}^{{\rm double}}}{{\mathcal N}^{{\rm single}}} ,
\label{eq:pgamma}
\ee
{where $\alpha_{S}^{(0)}\approx(1,1,0)$ represents the dominant single pion ($\pi^{+}$) contribution and $\alpha_{S}^{(1)}\approx(1,2,0)$ represents the contributions from double pion processes ($\pi^{+}$ and $\pi^{-}$).} 
{Based on the energy-loss time scales for each branching cross section~\cite{Murase:2010gj}}, the charged pion ratio can be estimated as 
\be
 \frac{{\mathcal N}^{{\rm double}}}{{\mathcal N}^{{\rm single}}} \approx \frac{\int d \bar{\varepsilon} \sigma^{{\rm double}}_{\pi^{-}} \int d\varepsilon \varepsilon^{-\beta-2}}{\int d \bar{\varepsilon} \sigma^{{\rm single}}_{\pi^{+}} \int d\varepsilon \varepsilon^{-\beta-2}}
\ee
where $\beta$ is the photon spectral index. Using the cross section data from Ref.~\cite{Schadmand:2005ji} we obtain the result shown in Fig.~\ref{figplusminus}. We shall use this to model the expected flavor ratio from a source in which photohadronic production of neutrinos is dominant. We also assume that the photon spectrum is soft enough, since multipion production becomes relevant for $\beta\lesssim1$~\cite{Murase:2005hy,Baerwald:2010fk}. Given the dependence on the charged pion ratio, it is in principle possible that one might be able to infer the photon spectral index from data, although this is quite challenging in reality (see, e.g., the bottom panel of Fig.~\ref{pgamma}).

Finally, we entertain the possibility that the relative abundance of $\pi^{-}$ and $\pi^{+}$ could be determine from flavor information for a $p\gamma$ source (see Eq.~\ref{eq:pgamma}). The resulting likelihood functions are shown in Fig.~\ref{pgamma} for a variety of input pion ratios. We see that the none of the likelihoods are very peaked, and that no strong conclusion can be reached for this variable. Were this to be possible one could infer the photon spectral index at the source (See. Fig~\ref{figplusminus}).

\vspace{2cm}

\bibliographystyle{JHEP}

\bibliography{nu}

\end{document}